\documentclass[aps,pra,reprint,groupedaddress]{revtex4-2}

\usepackage{graphicx}
\usepackage{amsmath,amssymb,multirow}
\usepackage{bm}
\usepackage{braket}
\usepackage[
colorlinks,
linkcolor=blue,
urlcolor=blue,
citecolor=blue
]{hyperref}

\begin{document}

\title{Autoionization of high-$\ell$ core-excited Rydberg states of alkaline-earth-metal atoms}

\author{E.\,Marin-Bujedo}
\author{M.\,G\'en\'evriez}
\email[]{matthieu.genevriez@uclouvain.be}
\affiliation{Institute of Condensed Matter and Nanosciences, Universit\'e catholique de Louvain, BE-1348 Louvain-la-Neuve, Belgium}

\date{\today}

\begin{abstract}

The autoionization of core-excited Rydberg states is theoretically studied for
a broad range of principal and angular-momentum quantum numbers $n$ and $\ell$
in alkaline-earth-metal atoms. We combined two theoretical methods to calculate
accurate autoionization rates for $n=10-65$ and $\ell=0-45$ over the
100 orders of magnitude that they span. The strong interaction between the two
valence electrons for low $\ell$ states is treated from first principles with
configuration interaction with exterior complex scaling, while at large $\ell$
the weak correlation is described by a perturbative approach and
arbitrary-precision floating-point arithmetics. The results, which we benchmark
against available experimental data, provide autoionization rates for the
$Np_{1/2, 3/2}$ and, when applicable, $(N-1)d_{3/2, 5/2}$ ion-core states of
Mg, Ca and Sr ($N=3-5$). Using the extensive set of calculated data, we analyze the
dependence of the rates on $\ell$ and identify five general laws of the
autoionization of high-$\ell$ states. An empirical formula describing the
scaling of the rates with $\ell$ is suggested.
\end{abstract}

\maketitle

\section{Introduction}

When the ion core of a Rydberg atom or molecule is excited the system can
decay via three different mechanisms: fluorescence of the ion core,
fluorescence of the Rydberg electron, and, because the energy of the system is
above the first ionization threshold, autoionization. Between these three
mechanisms, autoionization is the fastest by up to several orders of magnitude
for states in which the orbital angular momentum of the Rydberg electron is
low~\cite{aymar96}. The dynamics governing autoionization are a sensitive probe
to electron correlations and, as such, have been extensively studied both in the time and frequency
domain~\cite{camus89,warntjes99,wehrli19}. Experiments based on ion-core
excitation~\cite{cooke78a} have unraveled some of the fascinating electron
dynamics that take place in the dense manifolds of core-excited Rydberg
states~\cite{luc-koenig95,aymar96,pisharody04,eichmann92,jones90}, and studies
are ongoing to probe the complex correlations that occur for even higher
degrees of core excitation~\cite{genevriez23,fields18}. The development of
multi-channel quantum defect theory has led to a clear and powerful way of
understanding autoionization as the inelastic scattering of the Rydberg
electron off the ion core. An alternative method, the configuration interaction
with exterior complex scaling (CI-ECS), was recently
used~\cite{genevriez21,fields18,yoshida23} to describe the dynamics of
core-excited Rydberg states, in particular for higher-lying core excitation
where it provided a spectacular visualization of electron
dynamics~\cite{genevriez21a}.

The behavior of autoionization when the Rydberg electron has a low
orbital-angular-momentum quantum number $\ell$ has been extensively studied
(see, \textit{e.g.}, Refs.~\cite{aymar96,gallagher94} for reviews). In the
absence of series perturbations, the autoionization rates of a given series,
converging to a given ion-core state, scale with the principal quantum number
of the Rydberg electron as $n^{-3}$~\cite{gallagher94}. This scaling is no more than
the probability to find the Rydberg electron near the nucleus,
which is where the Rydberg electron inelastically scatters off the ion core and autoionizes. Autoionization for states with
high $\ell$ values, on the other hand, is much less well characterized. The centrifugal barrier
\begin{equation}
	\frac{\ell(\ell+1)}{2r^2}
\end{equation}
prevents the penetration of the Rydberg-electron wavefunction into the ion core
region, thereby suppressing autoionization. Pioneering experimental studies in
Sr have shown that the rates indeed drop rapidly with $\ell$~\cite{cooke78a}, a
result that was later verified for other series and
species~\cite{roussel90,lehec21,teixeira20,jones88} and confirmed by
theoretical predictions for $\ell \lesssim
10$~\cite{jones88,poirier88,poirier94a}. Fluorescence-decay mechanisms have
been observed to dominate the decay of core-excited Rydberg states for
sufficiently high $\ell$ values~\cite{wehrli19}. While values or upper limits
of the autoionization rates have been measured for $\ell$ as high as $50$~\cite{teixeira20},
theoretical values for $\ell \gtrsim 10$ are lacking, a fact that can be
attributed to the difficulty of calculating the matrix elements involved in the
rates.

The autoionization rates of high-$\ell$ core-excited Rydberg states play an
important role in pulsed-field-ionization zero-kinetic-energy photoelectron
spectroscopy~\cite{reiser88,merkt11}. Their low values stabilize core-excited
Rydberg states against autoionization~\cite{chupka93}, which permits the
measurement of photoelectron spectra of atoms, molecules and ions at high
resolution~\cite{hollenstein01,reiser88,wehrli21}. Autoionization also has
significant interest in cold-atoms experiments where it has been used to image
ultracold Rydberg gases~\cite{lochead13}, track the formation of ultracold
neutral plasmas~\cite{millen10} or realize high-fidelity state detection of
Rydberg atoms in an atomic array~\cite{madjarov20}. The possibility to
suppress autoionization offers many interesting properties for quantum optics
and quantum information experiments with Rydberg atoms~\cite{mukherjee11}.
Ion-core fluorescence, which can only be observed if it is faster than
autoionization, has been used to image ultracold Rydberg
gases~\cite{mcquillen13}, and optical control of the ion core is a promising
route to manipulate Rydberg atoms without perturbing the Rydberg
electron~\cite{muni22,pham22,burgers22}. In these perspectives, it appears
desirable to better understand the behavior of autoionization with $\ell$,
from regions where it predominates over other decay rates to regions where it
is completely suppressed.

We present a theoretical study of the autoionization of alkaline-earth-metal atoms
(Mg, Ca, Sr) in core-excited Rydberg states. These species were chosen for two
reasons. First, they are widely used in the quantum optics and quantum
information applications mentioned above. Second their electronic structure is
both amenable to high accuracy calculations and simple enough so that the
different dynamics governing autoionization can be identified and understood.

\begin{figure}
	\centering
	\includegraphics[width=0.95\columnwidth]{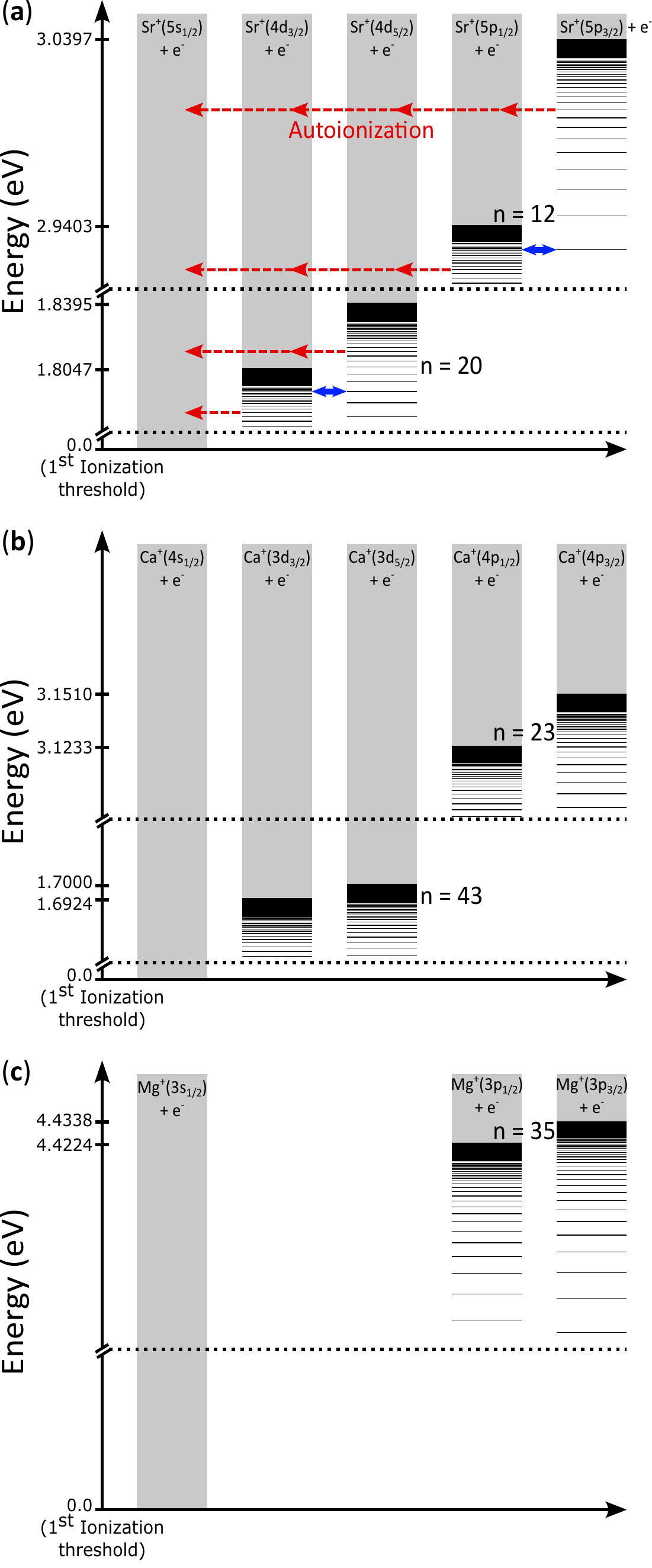}
	\caption{\textbf{(a)} Energy-level scheme of the relevant Rydberg states of Sr. The vertical scale is discontinuous. The red arrows represent autoionization processes. The blue double arrows illustrate other possible channel interactions between adjacent Rydberg series, responsible for series perturbations. For each ion-core fine-structure multiplet, we indicate the principal quantum numbers below which spin-orbit autoionization into the continua above the lowest fine-structure component is no longer possible. Same energy-level scheme for Ca \textbf{(b)} and Mg \textbf{(c)}.}
	\label{fig:Energy_schemes}
\end{figure}

We developed and used two theoretical methods to calculate autoionization
rates from $\ell=0$ all the way to $\ell=45$, for $n = 10 - 65$ and for
ion-core states comprising the excited states $Np_{1/2}$, $Np_{3/2}$, and, when
applicable, $(N-1)d_{3/2}$ and $(N-1)d_{5/2}$ ($N=3, 4$ and 5 for Mg, Ca and Sr, respectively). The relevant energy-level structures and
energy values of the three species are summarized in
Fig.~\ref{fig:Energy_schemes}. To calculate the rates over such a broad range
of states, we combined the capability of
CI-ECS~\cite{genevriez21,genevriez19b,genevriez21a} to treat the complete
two-electron dynamics from first principles with a perturbative treatment of
electron correlations to calculate the extremely small autoionization rates of
high-$\ell$ states with arbitrary numerical precision. The two methods are
discussed in detail in Sec.~\ref{sec:theory}. The results, presented in
Sec.~\ref{sec:results}, provide a complete picture of the autoionization rates of
the core-excited Rydberg states of Mg, Ca and Sr. They allow us to identify
general trends and properties of the autoionization of high-$\ell$ states,
which we rationalize by investigating the underlying electron dynamics. An empirical
formula describing the scaling of the rates with $\ell$ is suggested.

\section{Theory}\label{sec:theory}

\subsection{CI-ECS calculations}\label{sec:ci-ecs}

The description of core-excited Rydberg states is a challenging task for
atomic-structure techniques because it requires to treat the electronic motion
far from the nucleus ($r \sim 3000\ a_0$ for $n=45$), to calculate electronic
correlations over large regions of configuration space, and to describe
continuum processes and resonances. As in other studies (see
Ref.~\citenum{aymar96} for a review), we reduce the complexity of the problem
by treating alkaline-earth-metal atoms as quasi two-electron systems. The two
valence electrons, subject to the effective field of the closed-shell doubly
charged ion core, are considered explicitly. The effect of the remaining
electrons, on the other hand, is accounted for with a fitted effective core model
potential. The effective Hamiltonian describing the two valence electrons is
given by
\begin{align}
	\hat{H}(\bm{r_1},\bm{r_2}) =& -\frac12\bm{\nabla}_1^2 - \frac12\bm{\nabla}_2^2 + V_{\ell_1}(r_1) + V_{\ell_2}(r_2) + \frac{1}{r_{12}} \nonumber\\
	&+ V^\text{SO}_{\ell_1j_1}(r_1) + V^\text{SO}_{\ell_2 j_2}(r_2) + V^{(2)}_{\text{pol}}(\bm{r_1}, \bm{r_2})  ,
	\label{eq:two_electron_hamiltonian}
\end{align}
where the vectors $\bm{r_1}$ and $\bm{r_2}$ represent the positions of the two
electrons and $r_{12}$ is the distance between them. The Hamiltonian includes
$\ell$-dependent model potentials $V_{\ell_i}(r_i)$ representing the effect of the
doubly charged ion core on the valence electrons independently ($i=1, 2$). It also
includes the electron repulsion $1/r_{12}$ and the spin-orbit interaction
$V^\text{SO}_{\ell_i j_i}(r_i)$, with $j_i$ the total-angular-momentum quantum number of each electron. The two-electron term
$V^{(2)}_{\text{pol}}(\bm{r_1},
\bm{r_2})$ represents the polarization of the core upon the concerted motion
of the two electrons~\cite{genevriez21,hansen99,luc-koenig98}.

The model potentials $V_\ell(r)$ are of the form proposed in Ref.~\citenum{aymar96},
\begin{align}
	V_\ell(r) = &-\frac{1}{r}\left[ 2 + (Z-2)\text{e}^{-\alpha_1^\ell r} + \alpha_2^\ell \text{e}^{-\alpha_3^\ell r}\right]  \nonumber\\&- \frac{\alpha_{\text{cp}}}{2r^4}W_6\left(r; r_c^\ell\right)
		\label{eq:model_potential} ,
\end{align}
with the cutoff function $W_6$ defined as
\begin{equation}
	W_6\left(r; r_c^\ell\right) = 1 - \text{e}^{-(r/r_c^\ell)^6} .
	\label{eq:cutoff_function}
\end{equation}
The parameters $\alpha_1^\ell, \alpha_2^\ell, \alpha_3^\ell$ and $r_c^\ell$
have been optimized on the experimental values of the energy levels of the
singly-charged ion in Refs.~\cite{luc-koenig97}, \cite{aymar96}
and~\cite{luc-koenig98} for Mg$^+$, Ca$^+$ and Sr$^+$, respectively. Their
values are listed in Table~\ref{tab:modelpotential_parameters}.

The spin-orbit
interaction is given by~\cite{aymar96}
\begin{equation}
	V^{\text{SO}}_{\ell j}(r) = \alpha_\text{SO}^\ell \frac{\alpha^2}{2} \, \bm{\ell}\cdot\bm{s} \, \frac{1}{r} \frac{\text{d}V_\ell}{\text{d}r} \left[ 1 - \frac{\alpha^2}{2} V_\ell(r)\right]^{-2} ,
 	\label{eq:spin_orbit}
\end{equation}
with $\alpha$ the fine-structure constant. The additional scaling factor
$\alpha_\text{SO}^\ell$ was introduced and adjusted to reproduce the
spin-orbit splittings of the low-lying excited states of the ion with an
accuracy of better than $1$~cm$^{-1}$, instead of the 10~cm$^{-1}$ accuracy
obtained without it. Its values are also given in
Table~\ref{tab:modelpotential_parameters}. The 1-cm$^{-1}$ accuracy is
required to predict perturbations of the energies and autoionization rates
caused by Rydberg states of adjacent series with sufficient accuracy (see
Ref.~\citenum{genevriez19b} for examples with Mg). Without it, the perturbations would occur at the wrong energies and therefore for the wrong Rydberg states.

\begin{table}[ht]
\begin{ruledtabular}
\caption{Model-potential paramaters used in the calculations reported in this work.}
\label{tab:modelpotential_parameters}
\begin{tabular}{ccccccc}
	   & $\alpha_1^\ell$ & $\alpha_2^\ell$ & $\alpha_3^\ell$ & $r_c^\ell$ & $\alpha_\text{cp}$ & $\alpha_\text{SO}^\ell$\\\hline
	\multicolumn{7}{c}{Mg} \\
	$\ell=0$   & 4.51367 & 11.81954 & 2.97141 & 1.44776 & 0.49 & 1\\
	$\ell=1$   & 4.71475 & 10.71581 & 2.59888 & 1.71333 & 0.49 & 0.7875\\
	$\ell\ge2$ & 2.99158 & 7.69976  & 4.38828 & 1.73093 & 0.49 & 1\\
	\multicolumn{7}{c}{Ca} \\
	$\ell=0$   & 4.0616  & 13.4912  & 2.1539  & 1.5736 & 3.5 & 1\\
	$\ell=1$   & 5.3368  & 26.2477  & 2.8233  & 1.0290 & 3.5 & 0.984 \\
	$\ell=2$   & 5.5262  & 29.2059  & 2.9216  & 1.1717 & 3.5 & 0.68 \\
	$\ell\ge 3$   & 5.0687 & 24.3421& 6.2170  & 0.4072 & 3.5 & 1 \\
	\multicolumn{7}{c}{Sr} \\
	$\ell=0$   & 3.86849 & 7.89363 & 1.82951 & 1.11292 & 5.3 & 1\\
	$\ell=1$   & 3.43901 & 2.74445 & 1.48442 & 1.22661 & 5.3 & 0.982\\
	$\ell=2$   & 3.39035 & 4.32782 & 1.58635 & 1.55384 & 5.3 & 0.844\\
	$\ell\ge 3$   & 4.81077 & 4.06763& 1.75544& 0.94593 & 5.3 & 1
\end{tabular}
\end{ruledtabular}
\end{table}

The two-electron Schrödinger equation associated with the
Hamiltonian~\eqref{eq:two_electron_hamiltonian} is solved using the CI-ECS
method, which has been described in detail
elsewhere~\cite{genevriez19b,genevriez21,genevriez21a}. Briefly, the
two-electron wavefunction is written as a linear combination of
anti-symmetrized products of one-electron spin-orbitals. Angular momenta are
coupled in the $jj$ coupling scheme, which is the most appropriate for
core-excited Rydberg states~\cite{aymar96}. Autoionization and other continuum
processes are treated using the technique of exterior complex scaling
(ECS)~\cite{nicolaides78,simon79}. Following ECS, the radial coordinates $r_1$
and $r_2$ of the electrons are rotated into the complex plane by an angle
$\theta$ beyond a radius $R_0$,
\begin{equation}
	r \to \begin{cases}
		r & \qquad\text{if}\ r < R_0 \\
		R_0 + (r - R_0)\, \mathrm{e}^{\mathrm{i}\theta} &\qquad\text{if}\ r \ge R_0
	\end{cases} .
	\label{eq:ecs_contour}
\end{equation}

The interest of ECS lies in the behavior of resonance wavefunctions. For real
$r$ values, the amplitudes of resonance wavefunctions are nonnegligible even
as $r \to \infty$. Upon complex scaling, these become exponentially damped at
large distances and can be represented by square-integrable functions.
Calculations can thus be performed in a box of finite radius $r_\text{max}$
even when continua are involved. The size of the box limits the spatial extent of the largest Rydberg wavefunction that can be represented, and
therefore gives an upper bound to the maximal $n$ value that can be reliably
calculated. We typically choose $r_\text{max} > 10\,000\ a_0$ (see
Table~\ref{tab:femdvr_parameters}) such that $n_\text{max} \gtrsim 70$.

Complex scaling requires the use of complete square-integrable basis sets,
which would make the size of the two-electron basis set very large and the
calculations computationally demanding (see, \textit{e.g.},
Ref.~\cite{eiglsperger09}). This issue is overcome by choosing the
complex-rotation radius $R_0$ to be larger that the extension of the
core-electron wavefunction. In that case the core electron does not reside in
the complex-scaled region and is well described by a small number of radial
functions. Only the outer electron must be described by a (quasi)complete
basis set and the size of the two-electron basis set is dramatically reduced.

In practice, the one-electron spin-orbitals entering the two-electron
wavefunction are constructed from radial functions, spherical harmonics and
spinors (see~\cite{genevriez21} for details). The complex-scaled radial functions describing
each of the two electrons are numerical finite-element
discrete-variable-representation (FEM-DVR)
functions~\cite{rescigno00,genevriez19b}. They are obtained by solving the one-electron radial Schrödinger
equation for the singly-charged ion along the complex ECS
contour~\eqref{eq:ecs_contour}. In the FEM-DVR method, the radial space is split into several finite
elements $[r_i, r_{i+1}]$ and, in each element $i$, the Schrödinger equation is
solved on a grid of $N_i$ points with a Legendre-Gauss-Lobatto DVR
method~\cite{manolopoulos88}. We carefully chose the size of the finite
elements and the number of grid points to minimize the basis-set size and make
the calculations as fast as possible. The parameters of the FEM-DVR
calculations are listed in Table~\ref{tab:femdvr_parameters} for each
alkaline-earth-metal atom considered in this work.

In the CI expansion of the two-electron wavefunction, we use the quasi-complete
set of $1 + \sum_i (N_i - 1)$ FEM-DVR radial functions to describe the Rydberg
electron. The set of FEM-DVR functions representing the core electron is
restricted to those describing the low-lying levels of the Mg$^+$,
Ca$^+$ and Sr$^+$ ions listed in Table~\ref{tab:coreelectron_orbitals}. Together, this means that the
two-electron basis set comprises from $5\,000$ to $30\,000$ basis functions
depending on the total angular momentum, the parity, and the atomic species.

\begin{table}[t]
\begin{ruledtabular}
\caption{FEM-DVR paramaters used in the CI-ECS calculations reported in this work.}
\label{tab:femdvr_parameters}
\begin{tabular}{cccc}
	Element   & $[r_i, r_{i+1}]\ (a_0)$ & $N_i$ & $\theta_i (^\circ)$\\\hline
	\multicolumn{4}{c}{Mg} \\\hline
	$i=1$ & $[0, 100]$ & 100 & 0 \\
	$i=2$ & $[100, 17600]$ & 550 & 5 \\
	\multicolumn{4}{c}{Ca} \\\hline
	$i=1$ & $[0, 10]$ & 70 & 0 \\
	$i=2$ & $[10, 250]$ & 70 & 0 \\
	$i=3$ & $[250, 10250]$ & 300 & 5 \\
	\multicolumn{4}{c}{Sr} \\\hline
	$i=1$ & $[0, 10]$ & 80 & 0 \\
	$i=2$ & $[10, 150]$ & 80 & 0 \\
	$i=3$ & $[150, 13150]$ & 350 & 5 \\
\end{tabular}
\end{ruledtabular}
\end{table}

\begin{table}[t]
\begin{ruledtabular}
\caption{Core-electron orbitals included in the CI-ECS calculations reported in this work.}
\label{tab:coreelectron_orbitals}
\begin{tabular}{cl}
	Atom & Core-electron orbitals  \\\hline
	Mg & $3s_{1/2}, 3p_{1/2, 3/2}, 3d_{3/2, 5/2}, 4s_{1/2}, 4p_{1/2, 3/2}$ \\
	\multirow{2}{*}{Ca} & $4s_{1/2}, 3d_{3/2, 5/2}, 4p_{1/2, 3/2}, 5s_{1/2}, 4d_{3/2, 5/2}, 5p_{1/2, 3/2},$\\
	& $4f_{5/2, 7/2}, 5d_{3/2, 5/2}, 5f_{5/2, 7/2}$ \\
	\multirow{2}{*}{Sr} & $5s_{1/2}, 4d_{3/2, 5/2}, 5p_{1/2, 3/2}, 6s_{1/2}, 5d_{3/2, 5/2}, 6p_{1/2, 3/2},$\\
	&$4f_{5/2, 7/2}, 7s_{1/2}, 6d_{3/2, 5/2}, 7p_{1/2, 3/2}$ \\
\end{tabular}
\end{ruledtabular}
\end{table}

The Hamiltonian matrix~\eqref{eq:two_electron_hamiltonian} is calculated along
the ECS contour with the complex-scaled FEM-DVR basis and diagonalized. The eigenvalues and
eigenstates of the Hamiltonian are attributed, by inspecting the coefficients
of the CI expansion, to a Rydberg series with given values of $N$, $\ell_1$,
$j_1$, $\ell$, $j$ and $J$. The quantum numbers $N, \ell_1$ and $j_1$
correspond to the principal, orbital-angular-momentum and
total-angular-momentum quantum numbers of the core electron, respectively.
They indicate the ionization threshold to which the Rydberg series converges.
The quantum numbers $\ell$ and $j$ are associated to the angular momenta of
the Rydberg electron, and $J$ is the quantum number for the total angular momentum
of the entire, two-electron system. We use below the notation
$(N{\ell_1}_{j_1}n{\ell}_{j})_J$ to denote the Rydberg states. Because the
quantum defects $\delta_\ell$ of the high-$\ell$ states considered in this
work are very small, the difference between the principal quantum number $n$
of the Rydberg electron and its effective principal quantum $\nu = n -
\delta_\ell$ number is negligible in most cases. We thus use $n$
interchangeably to describe either quantity. When channel interactions are
strong, as is often the case for low $\ell$ values, the Rydberg series are
strongly perturbed and mixed, such that the assignment to one single series is
rather arbitrary.

\begin{figure}[ht]
	\includegraphics[width=\columnwidth]{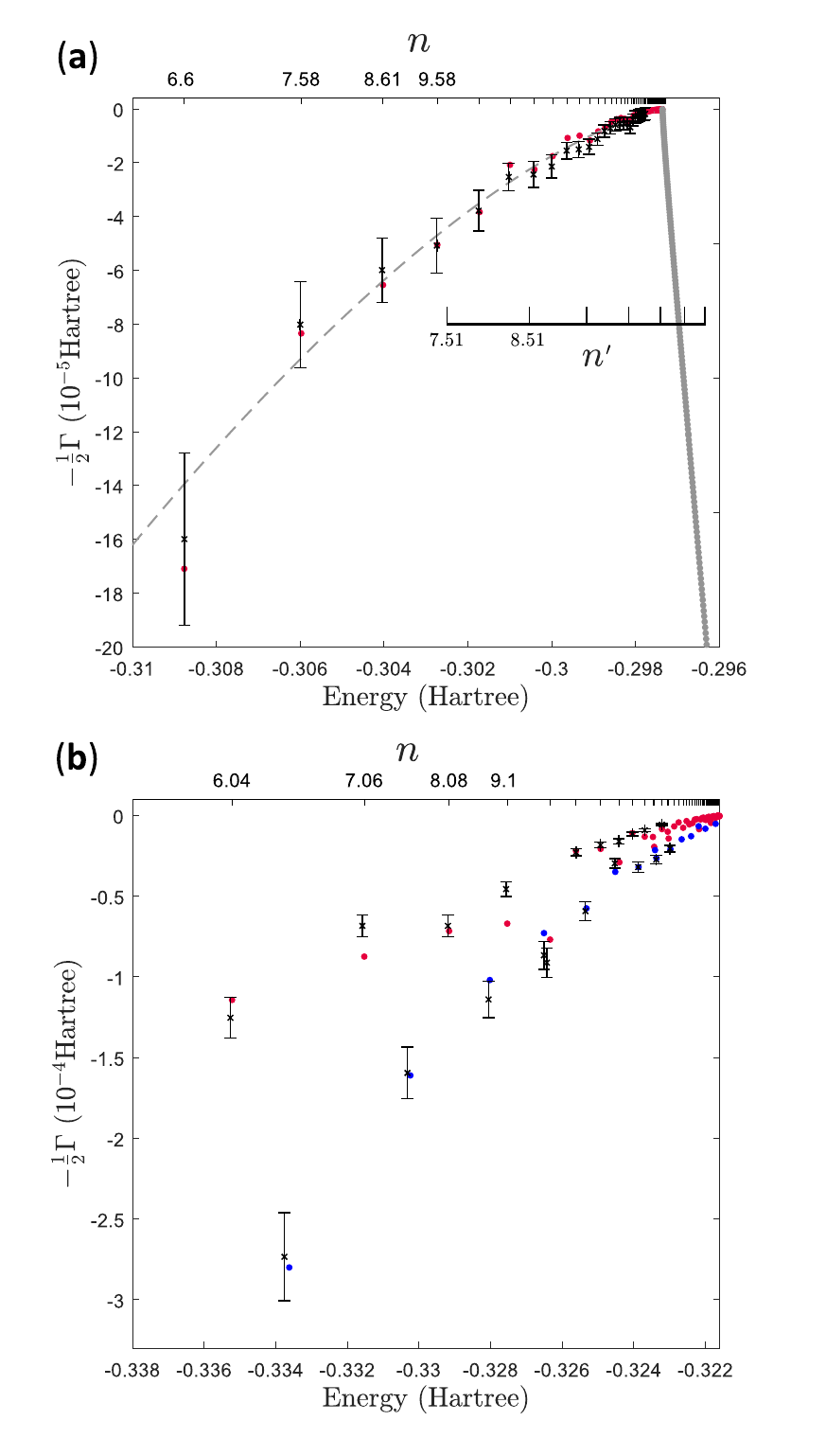}
	\caption{Comparison between the energies and autoionization rates of core-excited Rydberg states calculated with CI-ECS and measured experimentally for \textbf{(a)} the Sr$(5p_{1/2}ns_{1/2})_1$ Rydberg series~\cite{xu86}, and \textbf{(b)} the Ca$(4p_{1/2, 3/2}np_{j})_2$ series~\cite{bolovinos96}. The red and blue solid circles in panel \textbf{(b)} represent $4p_{1/2}$ and $4p_{3/2}$ ion core states, respectively. The effective principal quantum numbers $n$ relative to the $Np_{1/2}$ thresholds ($N=4,5$) are shown in the top horizontal axes. The assignment bar in panel \textbf{(a)} shows the effective principal quantum numbers $n'$ of Sr$(5p_{3/2}ns_{1/2})_1$ states. The gray dashed line shows the overall $n^{-3}$ scaling of autoionization rates.}
	\label{fig:ciecs_vs_experiment}
\end{figure}

Because the Hamiltonian matrix is
complex-symmetric, its eigenvalues are complex and given by $E -
\mathrm{i}\Gamma/2$ (see
Figs.~\ref{fig:ciecs_vs_experiment}(a) and ~\ref{fig:ciecs_vs_experiment}(b)
for the Sr$(5p_{1/2}ns_{1/2})_1$ and Ca$(4p_{1/2, 3/2}np_{j})_2$ series). When
the eigenstates correspond to bound states and autoionizing resonances (red
and blue solid circles), the eigenvalues are independent of the complex-rotation
angle $\theta$. The real part $E$ gives the energy of the state while the
imaginary part is half the autoionization rate $\Gamma$. The eigenvalues of
continuum states, on the other hand, are rotated with respect to the real axis
by $\sim 2\theta$ (gray solid circles).

To assess the accuracy and reliablity of our calculations, we have compared
the energies and autoionization rates of the calculated core-excited Rydberg
states against available experimental
data~\cite{xu86,bolovinos96,schinn91,dai90,xu87,assimopoulos94,jimoyiannis92}.
Overall, the agreement is excellent and the majority of the calculated rate
agree with experimental data within the uncertainties.

Two examples are shown in Fig.~\ref{fig:ciecs_vs_experiment} for the
Sr$(5p_{1/2}ns_{1/2})_1$ series (panel a) and Ca$(4p_{1/2, 3/2}np_{j})_2$
series (panel b), confirming the excellent agreement over the entire range of
$n$ values measured in the experiments. In the upper figure, perturbations
caused by the interaction of the Sr$(5p_{1/2}ns_{1/2})_1$ series with states
belonging to series converging to the Sr$^+(5p_{3/2})$ threshold cause
deviations from the smooth $n^{-3}$ decrease of the autoionization rates with
$n$ [dashed line in Fig.~\ref{fig:ciecs_vs_experiment}(a)]. The positions of
the perturber states, shown by the assignment bar within the figure, match the
energies at which the autoionization rates of $(5p_{1/2}ns_{1/2})_1$ states
are larger. This increase is caused by the mixing of these states with
the perturber, which has a lower $n$ value and therefore a larger
autoionization rate. In the lower figure, similar perturbations occur but the
larger number of Rydberg series involved makes the assignment of perturber
states more complicated.

The CI-ECS approach thus allows the accurate calculation of the energies and
autoionization rates of core-excited Rydberg states, even in regions where
perturbations between series are important. The extraction of the
autoionization rates from the calculations is straightforward and does not
involve fitting the density of states or the photoionization cross sections. It is
ideally suited for large-scale calculations of autoionization rates.

\subsection{Perturbation theory for high-$\ell$ states}\label{sec:lopt}

Although in principle the autoionization rates can be calculated with CI-ECS
for all values of $\ell$, this approach becomes cumbersome at high $\ell$
where the rates reach values below the numerical accuracy of the calculations
(typically $10^{-12}$ Hartree) and the numerical accuracy of double-precision
arithmetics on the computer ($10^{-16}$). Whereas abitrary-precision
arithmetics could be used to reach higher accuracies, they make the
calculation and diagonalization of the large complex-rotated Hamiltonian
matrix very demanding computationally. For large $\ell$ values the
centrifugal barrier experienced by the Rydberg electron is large and
prevents its penetration in the ion-core region. The interelectronic distance $r_{12}$
is always large and the electron repulsion is thus always small, such that a
full treatment of two-electron correlations is no longer necessary. Instead, a
perturbative treatment is possible which significantly simplifies the
calculations and makes the use of arbitrary-precision arithmetics possible.

In the perturbative limit, a core-excited Rydberg state and its associated
wavefunction are well described by a single $jj$-coupled configuration $(N
{\ell_1}_{j_1} n {\ell}_{j})_J$,
\begin{widetext}
\begin{equation}
	\ket{N\ell_1j_1n\ell jJM_J} =
	\sum_{\substack{m_{\ell_1}m_{\ell}\\m_{s_1}m_{s}\\m_{j_1}m_{j}}}
	\braket{\ell_1 m_{\ell_1} \frac{1}{2}m_{s_1} | j_1 m_{j_1}}\braket{\ell
	m_{\ell} \frac{1}{2}m_{s} | j m_{j}}\braket{j_1 m_{j_1}j m_{j} | J
	M_J}\ket{N\ell_1m_{\ell_1}\frac{1}{2}m_{s_1}}\ket{n\ell m_{\ell}\frac{1}{2}m_{s}}
	,
\end{equation}

\end{widetext}
with $\ket{n\ell m_\ell\frac{1}{2}m_s}$ and $\ket{N\ell_1m_{\ell_1}\frac{1}{2}m_{s_1}}$ describing the spin-orbitals of the Rydberg and core electrons, respectively. We omitted antisymmetrisation in the above because, for high-$\ell$ states, the effect of exchange is negligible as the
core and Rydberg electrons occupy very different regions of configuration space. The autoionization rate of a high-$\ell$ core-excited Rydberg state into a given continuum $(N' {\ell'_1}_{j'_1} \varepsilon {\ell'}_{j'})_J$ is given by Fermi's golden
rule,
\begin{equation}
	\Gamma = 2\pi \left|\braket{N'\ell'_1j'_1\varepsilon\ell'j'JM_J | \frac{1}{r_{12}} | N\ell_1j_1n\ell jJM_J}\right|^2 \rho(\varepsilon),
	\label{eq:golden_rule}
\end{equation}
with $\rho(\varepsilon)$ representing the continuum density of states at energy
$\varepsilon$.

The matrix element in Eq.~\eqref{eq:golden_rule} is calculated by expanding the electron-electron repulsion $1/r_{12}$ in multipole
terms~\cite{cowan81}. Although there is a large number of terms in the expansion, we have observed that only the dipole ($q=1$) and quadrupole ($q=2$) terms contribute significantly
to the calculated rates, \textit{i.e.}, $\Gamma \simeq \Gamma^{(1)} + \Gamma^{(2)}$. For each multipole term $\Gamma^{(q)}$, carrying
out the integration over all coordinates gives
\begin{equation}
	\Gamma^{(q)} = 2\pi \left[R_{N \ell_1 j_1}^{N' \ell'_1 j'_1, q}R_{n \ell}^{\varepsilon \ell', q}\right]^2 \left|B^{(q)}\right|^2 ,
	\label{eq:lopt_rate}
\end{equation}
with the squared norm of the angular integral $B^{(q)}$ given by
\begin{align}
	\left|B^{(q)}\right|^2 =& [j_1, j_1', j, j', \ell_1, \ell_1', \ell, \ell']
	\nonumber\\
	& \times 
	\begin{pmatrix}
		\ell_1' & q & \ell_1 \\
		0    & 0 & 0
	\end{pmatrix}^2
	\begin{pmatrix}
		\ell' & q & \ell \\
		0    & 0 & 0
	\end{pmatrix}^2
	\begin{Bmatrix}
		j_1  & q & j'_1 \\
		\ell_1' & 1/2 & \ell_1
	\end{Bmatrix}^2 \nonumber\\
	&\times
	\begin{Bmatrix}
		j  & q & j' \\
		\ell' & 1/2 & \ell
	\end{Bmatrix}^2
	\begin{Bmatrix}
		j'  & j_1' & J \\
		j_1 & j & q
	\end{Bmatrix}^2 .
	\label{eq:angular_integral_jj}
\end{align}
We used the usual notation $[a, b, \ldots] = (2a+1)(2b+1)\ldots$. 

The radial integral
\begin{equation}
	R_{N \ell_1 j_1}^{N' \ell'_1 j'_1, q} = \int \mathrm{d}r_1\, u_{N'\ell'_1j'_1}(r_1) r_1^q u_{N\ell_1j_1}(r_1)
\end{equation}
involving the reduced radial wavefunctions $u_{n\ell j}(r)$ of the core
electron is calculated by considering that the influence of the distant
Rydberg electron on the core electron is minimal, and that the core-electron
radial wavefunction is identical to the one of the bare ion. The integral is
then calculated using the ionic FEM-DVR basis functions obtained, as for the
CI-ECS calculations, by solving the one-electron Schrödinger equation for the
ion (see Sec.~\ref{sec:ci-ecs}).

The radial integral for the Rydberg electron,
\begin{equation}
	R_{n \ell}^{\varepsilon \ell', q} = \int \mathrm{d}r_2\, u_{\varepsilon\ell'}(r_2) r_2^{-q - 1} u_{n\ell}(r_2) ,
\end{equation}
is calculated using the fact that the Rydberg electron in a high-$\ell$ state experiences, to a very good approximation, the Coulomb potential of the singly charged ion core and nothing else. Because of the large centrifugal barrier, it does not penetrate into the core region where the electrostatic potential would depart from the Coulomb case. In other words, the quantum defect is vanishingly small (see~\cite{gallagher94} and~\cite{aymar96} for details) and the Rydberg-electron radial wavefunction is hydrogenic. The integral is then known analytically in terms of the Appell hypergeometric function $F_2$~\cite{matsumoto91},
\begin{widetext}
\begin{align}
	R_{n \ell}^{\varepsilon \ell', q} = & \mathcal{N}_{n,\ell}\mathcal{C}_{\varepsilon, \ell'}\Gamma(\ell + \ell' - q + 2)\left(\frac{1}{n} + \mathrm{i}k\right)^{-(\ell + \ell' - q + 2)}
	\nonumber\\
	& \times F_2(\ell + \ell' - q + 2, -\frac{1}{\mathrm{i}k}+\ell'+1, 2\ell+2, 2\ell'+2; \frac{2}{1+\mathrm{i}kn}, \frac{2\mathrm{i}kn}{1+\mathrm{i}kn}) ,
	\label{eq:radial_rydberg}
\end{align}
\end{widetext}
with $k = \sqrt{2\varepsilon}$. The normalization constant $\mathcal{N}_{n,\ell}$ of the initial state is given by
\begin{equation}
	\mathcal{N}_{n,\ell} = \frac{1}{(2\ell+1)!}\sqrt{\frac{(n+\ell)!}{(n-\ell-1)!2n}}\left(\frac{2}{n}\right)^{\ell+3/2}
\end{equation}
and the one of the final state by
\begin{equation}
	\mathcal{C}_{\varepsilon, \ell} = \frac{1}{(2\ell+1)!}\frac{2(2k)^\ell}{\sqrt{1-\exp(-2\pi/k)}}
	\prod_{s=1}^\ell \sqrt{s^2 + \frac{1}{k^2}} .
\end{equation}

The Appell $F_2$ function, along with all other functions in
Eq.~\eqref{eq:radial_rydberg}, can be calculated to within arbitrary numerical
precision with the mpmath library~\cite{thempmathdevelopmentteam23}. For the
calculations presented below, a numerical accuracy of $10^{-60}$ was chosen for
the calculation of the radial integrals~\eqref{eq:radial_rydberg}, which have values ranging from about
$1$ atomic unit at low $\ell$ to $10^{-50}$ atomic units at high $\ell$. The
other quantities entering Eq.~\eqref{eq:lopt_rate} have values well above the
numerical precision of the computer and can be calculated using
double-precision arithmetics. The final result is obtained, using a numerical
accuracy of $10^{-120}$, by multiplying and squaring all quantities together to
obtain $\Gamma^{(q)}$, by summing over the dipole and quadrupole contributions,
and by summing over all continua accessible from the core-excited Rydberg state
under consideration. Because the squares of the angular integrals have values
of typically $10^{-2}$ or larger, the relative numerical accuracy of the final
results is guaranteed to be at least $10^{-10}$ for values up to $10^{-110}$
atomic units. This means that all the rates shown below, whose values reach as
low as $10^{-100}$, are calculated with sufficient numerical accuracy.
Determining such minuscule autoionization rates is possible only with
theoretical methods and not with experimental measurements, because the
lifetimes involved are far too long and other decay mechanisms will dominate.

Anticipating on the results presented below, the rates calculated with the
perturbative approach closely match the ones obtained by solving the full
two-electron Schrödinger equation with CI-ECS for $\ell$ values in the range
from 6 to 10 (see for example the blue and red circles in
Fig.~\ref{fig:Partial_rates_Sr}). Above $\ell \sim 10$, the rates are in
general lower than the numerical accuracy of the CI-ECS calculations and only
the perturbative approach provides reliable results. Below $\ell \sim 6$,
perturbations are frequent. Because they cannot be represented within the
single-configuration framework of the perturbative approach, only the CI-ECS
method provides reliable results in this range. The agreement between the two
methods in the $\ell \sim 6-10$ range validates the perturbative treatment for
large $\ell$ values and shows that it is possible, when combining the two
approaches, to accurately calculate the autoionization rates of core-excited
Rydberg states over the \emph{entire} range of possible $\ell$ values. The
results provide benchmark data for future studies and, because they permit a
systematic analysis of the role played by the Rydberg-electron angular
momentum, they allow us to gain deep physical insight on the electronic
dynamics responsible for autoionization.

\section{Results}\label{sec:results}

We calculated the autoionization rates of all states of Mg, Ca, and Sr with $10
\le n \le 65$ and $0 \le \ell \le 45$ with the methods described above. The
numerical results are provided in the Supplemental Material. In the following,
we analyze the $\ell$ dependence of the rates and extract general behaviors
from the large body of calculated data. In most cases, the rates decrease
rapidly with $\ell$, as expected from previous
works~\cite{cooke78a,roussel90,lehec21,teixeira20,jones88,poirier88,poirier94a},
but do not always follow a single decay trend. For $\ell \gtrsim 10$, the
autoionization rates differ by several orders of magnitude depending on the
values of $j$ and $J$. For a given ion-core state and fixed values of $j - \ell$ and $J-\ell$, the
evolution of the rates with $\ell$ is smooth and, anticipating on the results
of Sec.~\ref{sec:scaling}, follows simple scaling laws. We will see that this
behavior is in fact governed by the value of $K- \ell$, with $K$ the quantum
number associated with the total angular momentum without Rydberg-electron
spin. This allows us to define \emph{branches} as ensembles of Rydberg states converging to a given ionization threshold and 
with fixed values of $K - \ell$, whose autoionization rates behave
in a similar manner. Such branches can exhibit a fine structure due to the
spin-orbit interaction of the Rydberg electron. We first analyze the behavior
of single branches, before considering all branches and later all thresholds of
all species.

\subsection{Behavior for a single branch}

\begin{figure}
    \centering
	\includegraphics[width=\columnwidth]{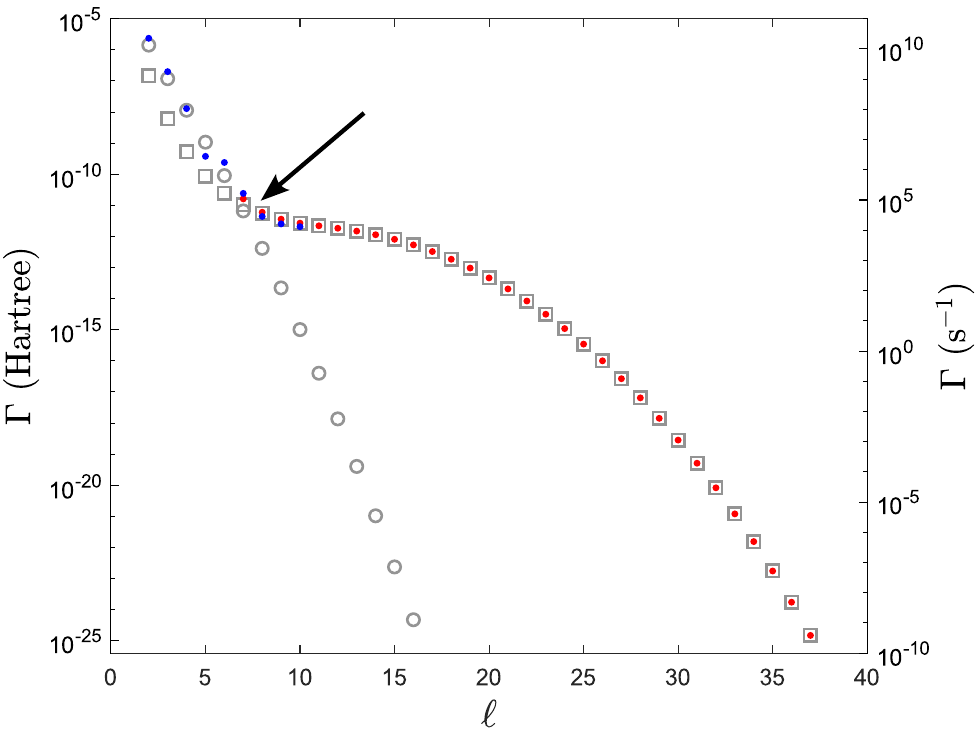}
    \caption{Autoionization rates of the $(4d_{5/2} 45{\ell}_{j})_J$ core-excited Rydberg states of Sr as a function of the orbital quantum number of the Rydberg electron ($j = \ell-1/2$ and $J = j + 1/2$). Solid blue circles represent the total rates calculated for low $\ell$ values with the CI-ECS method (Sec.~\ref{sec:ci-ecs}) and red ones show the total rates calculated for large $\ell$ values with the perturbative approach of Sec.~\ref{sec:lopt}. The empty gray symbols show the partial decay rates to continua above the $5s_{1/2}$ (circles) and $4d_{3/2}$ (squares) ionization thresholds calculated with the perturbative approach.}
    \label{fig:Partial_rates_Sr}
\end{figure}

Figure~\ref{fig:Partial_rates_Sr} shows the autoionization rates of the $\left(
4d_{5/2}45\ell_{j} \right)_J$ core-excited Rydberg states of Sr with $j =
\ell - 1/2$ and $J = j +1/2$ ($K-\ell=+\frac{1}{2}$). The calculated rates decrease
by more than 20 orders of magnitude between $\ell = 1$ and $\ell = 44$. For
$\ell \gtrsim 25$, they are far smaller ($\Gamma \lesssim 1$~s$^{-1}$) than those
of other decay mechanisms such as the fluorescence of the Rydberg electron,
which takes place in the milliseconds range.

The decay of the rates in Fig \ref{fig:Partial_rates_Sr} does not follow a
single trend and a shoulder (shown by the arrow) is observed around $\ell = 7$. Its
origin is, predominantly, the vastly different behavior of autoionization into
the continua above the Sr$^+ \left( 5s_{1/2} \right)$ and Sr$^+ \left( 4d_{3/2}
\right)$ ionization thresholds, both accessible from the $\left(
4d_{5/2}45\ell_{j} \right)_J$ states (see Fig.~\ref{fig:Energy_schemes}).
For the former threshold, the partial rates (empty gray circles) are large for
small $\ell$ and fall very rapidly as $\ell$ increases. For the latter
threshold, the partial rates (empty gray squares) are significantly smaller
for small $\ell$ but decrease more slowly with $\ell$ and thus dominate
the total autoionization rates for $\ell \gtrsim 10$.

Autoionization into the continua above both the Sr$^+(5s_{1/2})$ and
Sr$^+(4d_{3/2})$ thresholds is predominantly caused by the quadrupolar part of
the electron-electron repulsion. The kinetic energy of the emitted electron is
however much larger for the $5s_{1/2}$ continua ($1.84$~eV) than for the
$4d_{3/2}$ ones ($0.035$~eV). For large kinetic energies, the radial
integral~\eqref{eq:radial_rydberg} for the Rydberg electron decreases much
faster with $\ell$ than for smaller kinetic energies, a fact we verified in
a systematic manner for electron kinetic energies from 0.3 eV to 8 eV and
all possible values of $\ell$. The other quantities entering the
autoionization rates given by Eq.~\eqref{eq:lopt_rate} vary only little or not
at all with $\ell$. Therefore, because of the Rydberg-electron radial
integral, the rates decrease faster with $\ell$ for larger electron kinetic
energies. We verified this property for the other series and thresholds of Sr,
Ca and Mg. In conclusion, the shoulder in the rates of Fig
\ref{fig:Partial_rates_Sr} comes from the fact that two continua with very different energies are accesible upon autoionization.

Another shoulder can be observed for the partial rates into the continua above
the Sr$^+(4d_{3/2})$ threshold (empty gray squares). In this case, it cannot
be attributed to different photoelectron kinetic energies. Instead, it is
explained by a change of the values of the angular integrals in
Eq.~\eqref{eq:lopt_rate} which finds its origin in the
evolution of the angular-momentum coupling between the core and Rydberg
electrons discussed in the following section.

\subsection{Behavior for all branches}\label{sec:results_allbranches}

\begin{figure}
    \centering
	\includegraphics[width=\columnwidth]{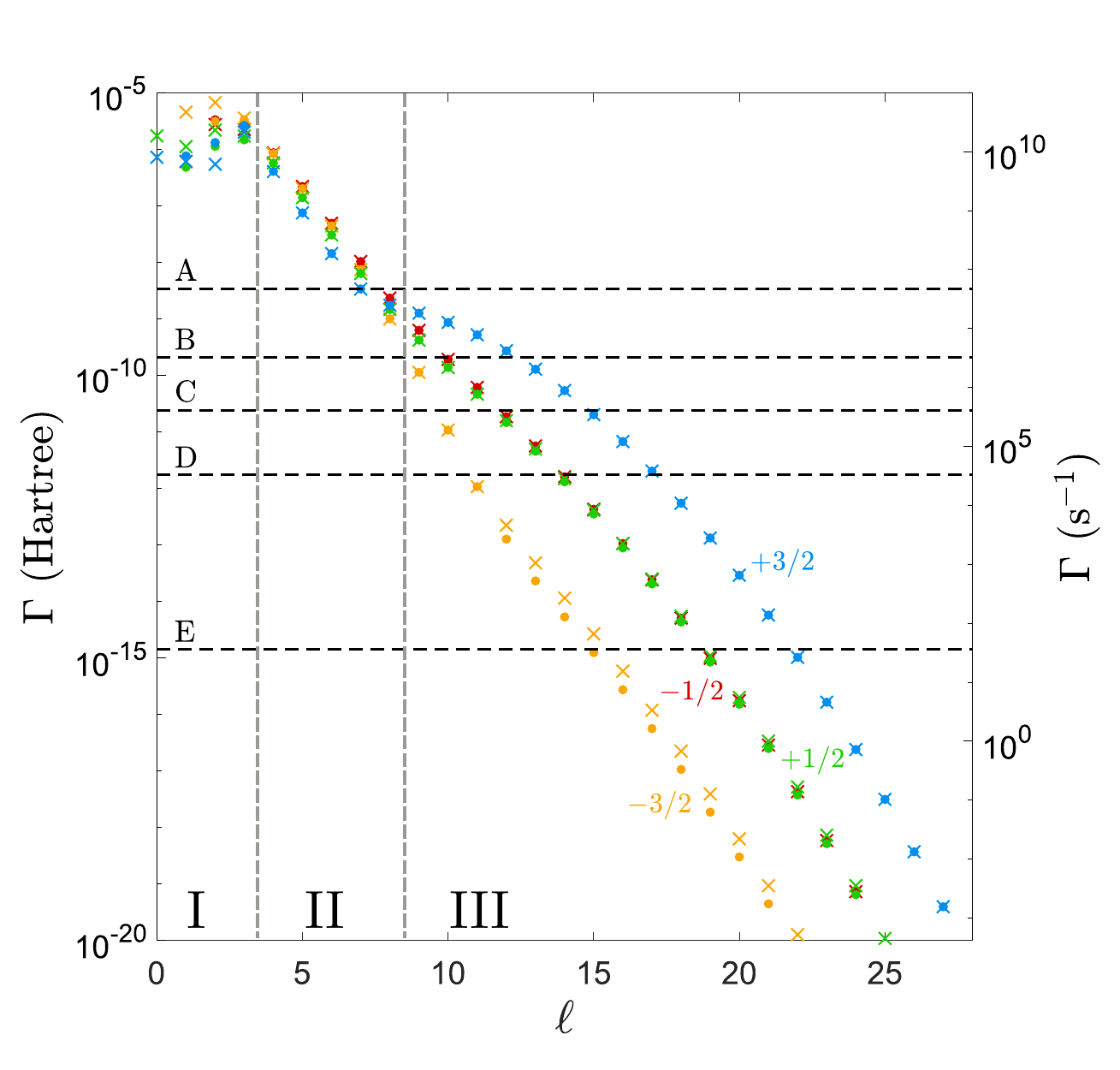}
    \caption{Autoionization rates of the $(5p_{3/2}45{\ell}_{j})_J$ states of Sr as a function of $\ell$ for all possible $j$ and $J$ values. The different branches are labeled, and their color is chosen, according to the value of $K - \ell$ (see text). Each branch has a fine structure corresponding to the fine structure of the Rydberg electron. The two fine-structure components are associated with the two possible values of the coupling between $K$ and the spin of the Rydberg electron (solid circles for $J = K - \frac{1}{2}$ and crosses for $K+\frac{1}{2}$). The horizontal dashed lines labeled A, B, C, D and E represent the decay rates of the radiative deexcitation channels $5p_{3/2} \rightarrow 5s_{1/2}$, $5p_{3/2} \rightarrow 4d_{5/2}$, $5p_{3/2} \rightarrow 4d_{3/2}$, $5s45p \rightarrow 5s^2$ and $45 \,c \rightarrow 44\,c$, respectively, with $c$ representing circular Rydberg states.}
    \label{fig:Branches_Sr(5p32)}
\end{figure}

We now consider the behavior of autoionization rates with $\ell$ for all
possible values of $j$ ($\ell \pm \frac{1}{2}$) and $J$ ($j + j_1 \ge J
\ge |j - j_1|$). The value of $n$ is fixed and we consider a single
ionization threshold ($N, \ell_1, j_1$ fixed). Figure
\ref{fig:Branches_Sr(5p32)} shows the rates of the
$(5p_{3/2}45\ell_{j})_J$ states of Sr. At low $\ell$ values,
the Rydberg electron penetrates into the ion core region. Its wave function
significantly overlaps with the one of the core electron and the quantum
defects are large. Perturbations between adjacent Rydberg series are frequent
(see Fig~\ref{fig:ciecs_vs_experiment}(b) for example) and the autoionization
rates do not exhibit a regular behavior for $\ell \lesssim 4$ (region I in
Fig. \ref{fig:Branches_Sr(5p32)}). In an intermediate region between $\ell
\sim 4$ and $\ell \sim 8$ (region II), the rates decrease monotonically and
their magnitudes are similar for all values of $j$ and $J$. In the last
region (region III), they split into what appears to be three branches with
very different magnitudes but similar evolution with $\ell$.
The apparent branches are further split by the spin-orbit interaction of the Rydberg electron, leading to two fine-structure components with slightly different rates (solid circles and crosses in Fig \ref{fig:Branches_Sr(5p32)}).

The branches can be associated to different values of $K-\ell$, $K$
being the quantum number of the total angular momentum excluding
Rydberg-electron spin. This is not surprising because, when the spin-orbit
interaction of the Rydberg electron is negligible, intermediate ($jK$)
coupling is the most appropriate coupling scheme for core-excited Rydberg
states~\cite{aymar96}. The total angular momentum $\bm{j}_1$ of the core
electron strongly couples to the orbital angular momentum $\bm{\ell}$ of the
Rydberg electron to give $\bm{K}$. The $jj$-coupled states that we calculate
with the methods of Sec.~\ref{sec:theory} are related to intermediate-coupling
states by the transformation coefficients~\cite{cowan81}
\begin{align}
	\ket{N\ell_1j_1n\ell jJ} = \sum_{K}& (-1)^{j_1 + \ell + \frac{1}{2} + J} \sqrt{(2K+1)(2j + 1)}\nonumber\\
	&\begin{Bmatrix}
		j_1 & \ell & K \\
		\frac{1}{2} & J & j
	\end{Bmatrix}
	\ket{N\ell_1j_1n\ell KJ} .
\end{align}
For large $\ell$ values, there is an almost one-to-one correspondence between
a given $jK$-coupled state, which we denote as $(N{\ell_1}_{j_1}n\ell)_K$
below, and the two $jj$-coupled $(N{\ell_1}_{j_1}n{\ell}_{j=\ell \pm
1/2})_{J=K \pm 1/2}$ states, thus making the assignment straightforward.

For the rates shown in Fig.~\ref{fig:Branches_Sr(5p32)}, there are four
possible values of $K$ for each value of $\ell$ ($K - \ell = \pm
\frac{1}{2}, \pm \frac{3}{2}$). The four corresponding branches are
distinguished by their color and labelled by their $K - \ell$ value. The
spin-orbit interaction of the Rydberg electron further splits the $K-\ell$
branches into two sub-branches with $J = K + 1/2$ and $J = K - 1/2$ (crosses
and solid circles in Fig.~\ref{fig:Branches_Sr(5p32)}, respectively). This
effect is very small, as expected because the spin-orbit interaction of the
Rydberg electron is small, and it is in fact only visible for the $K - \ell = -
3/2 $ branch (orange crosses and circles).

\begin{figure}[h]
    \centering
	\includegraphics[width=\columnwidth]{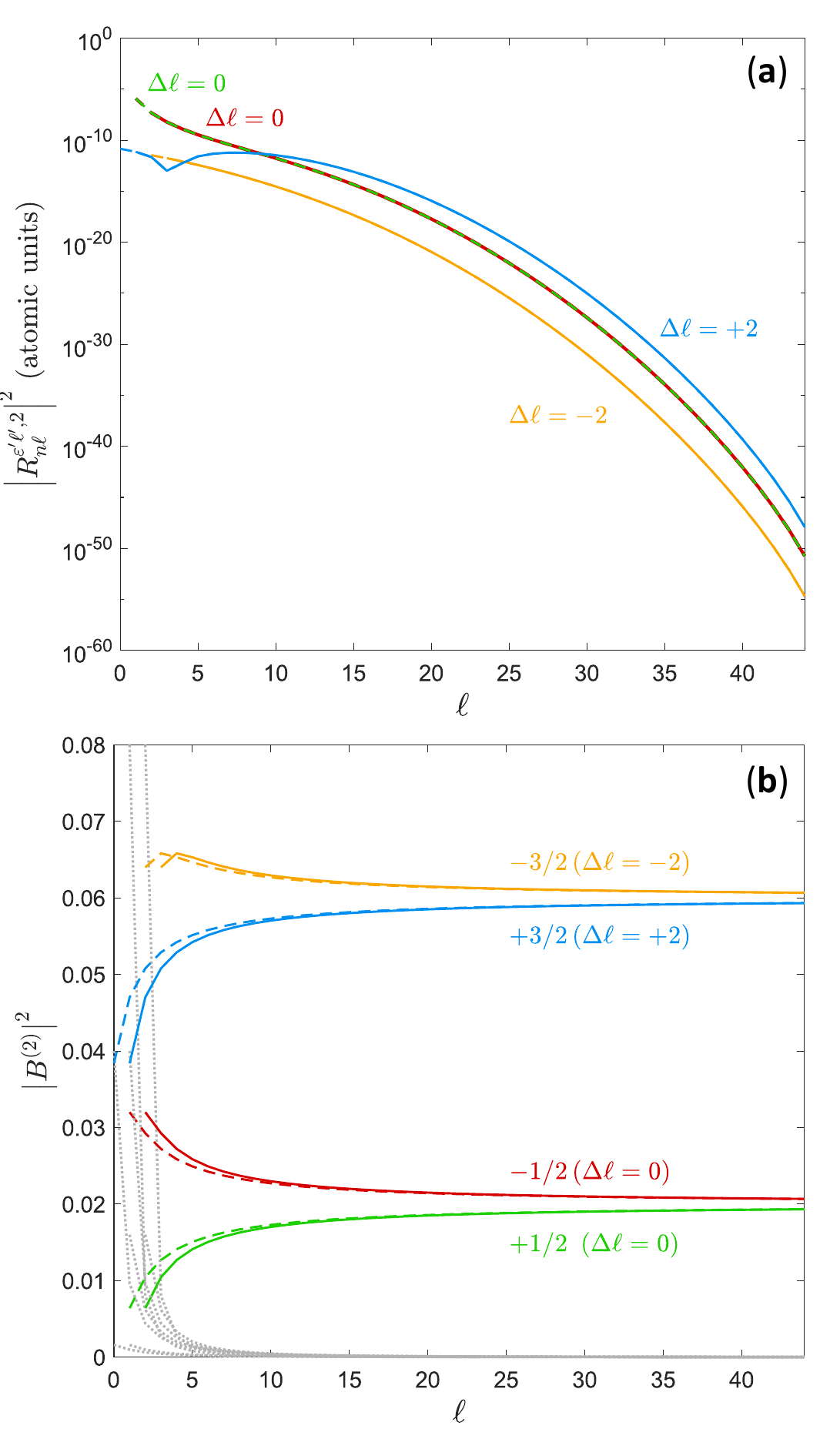}
    \caption{$\ell$ dependence of the norm squared of \textbf{(a)} the radial integrals for the Rydberg electron~\eqref{eq:radial_rydberg} and \textbf{(b)} the angular integrals~\eqref{eq:angular_integral_jj} for Sr$(5p_{3/2}45{\ell}_{j})_J$  states with all possible $j$ and $J$ values. Orange, red, green and blue lines correspond to $K-\ell = -3/2$, $-1/2$, $+1/2$ and $+3/2$ branches, respectively. The solid and dashed lines for the angular integrals represent the two possible couplings of the Rydberg-electron spin with $K$ ($J=K-1/2$ and $+1/2$, respectively). The gray lines are angular integrals for which $J$ is conserved but not $K$. For the colored lines, both $J$ and $K$ are conserved}
    \label{fig:wavefunctions}
\end{figure}

The very different magnitudes of the autoionization rates of the branches
trace back to the interplay between the radial and angular parts of the
electron-electron repulsion in Eq.~\eqref{eq:lopt_rate}. For $\ell \gtrsim
10$, autoionization into the continua above the $5p_{1/2}$ threshold dominates
over autoionization into those above the $4d_{3/2, 5/2}$ and $5s_{1/2}$
thresholds because the photoelectron kinetic energy is much smaller in the
former case (see Fig.~\ref{fig:Energy_schemes}). The
$(5p_{3/2}45{\ell}_{j})_J$ states couple to the $5p_{1/2}$ continua
through the quadrupolar (tensor order of $2$) part of the electron-electron repulsion ($q = 2$).
Upon autoionization, $\ell$ is thus unchanged or changes by $2$ ($\Delta
\ell = \ell' - \ell = 0,\, \pm2$). Importantly, the radial integrals
$\langle
\varepsilon \ell'|1/r_2^3|n\ell \rangle$ differ by several orders of magnitude depending on the value
of $\Delta \ell$, with $\Delta \ell = + 2$ being the largest [see
Fig.~\ref{fig:wavefunctions}(a)]. The same observation applies for the dipole (tensor order of 1)
part of the electron repulsion ($q=1$), in which case the radial integrals are
much larger for $\Delta \ell = +1$ than for $\Delta \ell = -1$. A similar
situation is encountered for the dipole matrix elements describing
photoabsorption and photoionization in hydrogenic systems, for which $\Delta
\ell = + 1$ transitions dominate over $\Delta \ell = -1$
transitions~\cite{bethe57}.

Because $K$ (or $J$ for $jj$ coupling) must be conserved upon autoionization,
angular-momentum coupling constrains the possible changes of $\ell$ for a
given branch. Considering the quadrupolar interaction, the initial state
$(5p_{3/2}n\ell)_{K = \ell - 3/2}$ can only autoionize into
continua above the $5p_{1/2}$ threshold with $\ell' = \ell - 2$. Continua with $\ell' = \ell + 0 ,\, 2$ are inaccessible because in these
cases angular-momentum coupling between the $5p_{1/2}$ core electron and the
$\ell'$ ionized electron cannot yield $K=\ell - 3/2$ and $K$ cannot be conserved. The autoionization of the states of the $- 3/2$ branch thus involves a $\Delta \ell = -2$ transition only, with a small radial integral translating into small values for the rates (orange crosses and
circles in Fig.~\ref{fig:Branches_Sr(5p32)}). The opposite observation holds for
the $K - \ell = +3/2$ branch. Only the $\Delta
\ell = +2$ transition is possible and, because the corresponding radial
integral is large, the autoionization rate is large (blue crosses and
circles in Fig.~\ref{fig:Branches_Sr(5p32)}). For the $K - \ell = \pm 1/2$ branches (red and green circles and
crosses in Fig.~\ref{fig:Branches_Sr(5p32)}), the only possible transition is $\Delta \ell = 0$, which explains
why the rates of both branches are very similar and lie between those of the
$+3/2$ ($\Delta \ell = +2$) and  $-3/2$ ($\Delta \ell = -2$) branches.

\begin{figure}
    \centering
	\includegraphics[width=\columnwidth]{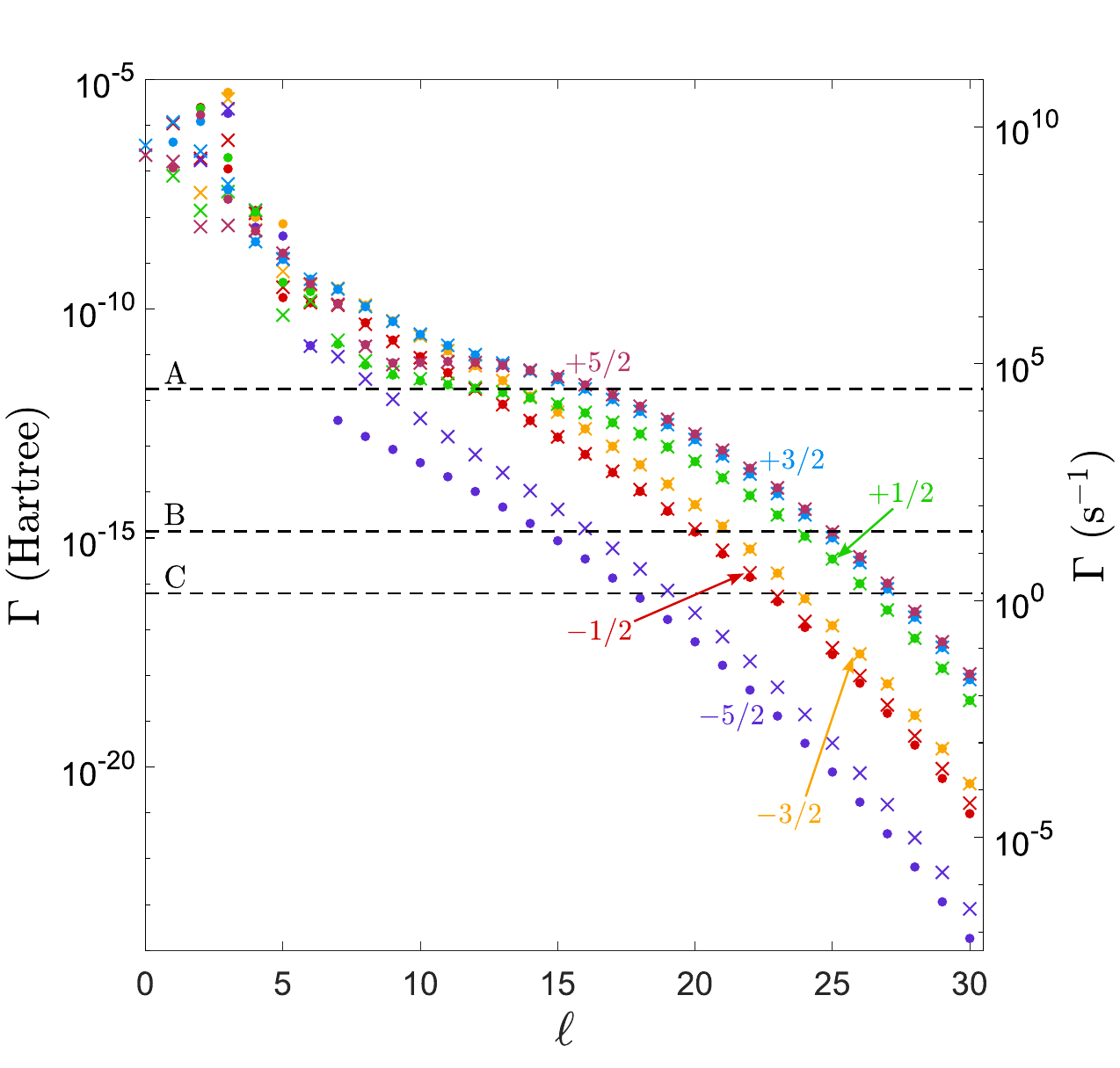}
    \caption{Autoionization rates of the $(4d_{5/2}45{\ell}_{j})_J$ states of Sr as a function of $\ell$ for all possible $j$ and $J$ values. The symbols and colors are defined as in Fig.~\ref{fig:Branches_Sr(5p32)}. The horizontal dashed lines labeled A, B and C represent the decay rates of the radiative deexcitation channels $5s45p \rightarrow 5s^2$, $45 \,c \rightarrow 44\,c$, and $4d_{5/2} \rightarrow 5s_{1/2}$, respectively, with $c$ representing circular Rydberg states.}
    \label{fig:Branches_Sr(4d52)}
\end{figure}

The above analysis revealed that the gross structure of the branches is
related to which $\Delta \ell$ value contributes predominantly to
autoionization. Let us consider another example, the $4d_{5/2}n\ell$
states of Sr, which for large $\ell$ values autoionize predominantly into
continuum above the Sr$^+(4d_{3/2})$ threshold through quadrupole
interactions. As illustrated in Fig.~\ref{fig:Branches_Sr(4d52)}, six branches
($K - \ell =\pm\frac{1}{2}, \pm\frac{3}{2}$ and $\pm\frac{5}{2}$) can be observed which
are grouped into 3 main components. The lowest branch and component, which is
further split by the spin-orbit interaction of the Rydberg electron, is $K -
\ell = -5/2$ and autoionizes through $\Delta \ell = -2$ transitions only.
The intermediate branches $K - \ell = -3/2$ and $-1/2$ autoionize through
predominantly $\Delta \ell = 0$ transitions, and the branches with the
largest rates, $K - \ell = 1/2, 3/2$ and $5/2$ autoionize through strong
$\Delta \ell = +2$ transitions. The substructure within the three main
$\Delta \ell$ components is due to differences in angular-momentum coupling
which translate into different values of the angular integrals
in Eq.~\eqref{eq:lopt_rate}.

The conclusions drawn above implicitly rely on the assumption that the angular
integrals entering the autoionization rates have similar magnitudes for any
$\Delta \ell$. As shown in Fig.~\ref{fig:wavefunctions} this is verified for
large $\ell$ ($\ell \gtrsim 10$) and thus $K$ values. The
underlying reason can be made explicit by considering the large-$\ell$
behavior of the angular integrals. We use $jK$ coupling to make the behavior
more apparent, however similar conclusion can be drawn with $jj$ coupling as
well. As shown in appendix~\ref{sec:asymp_jk_coupling}, the asymptotic
formulas for the Wigner symbols~\cite{varshalovich88b} allow to reduce the norm squared
of the angular integral $B^{(q)}$ in $jK$ coupling to
\begin{align}
\left| B^{(q)} \right|^2 \sim & [\ell_1, \ell_1', j_1]
    \begin{pmatrix}
		\ell_1' & q & \ell_1 \\
		0    & 0 & 0
	\end{pmatrix}^2
	\begin{Bmatrix}
		j_1  & q & j'_1 \\
		\ell_1' & 1/2 & \ell_1
	\end{Bmatrix}^2
	\nonumber\\
	& \times
	\left[D_{0\Delta \ell}^q(0, \frac{\pi}{2}, 0)\right]^2
	\nonumber\\
	&\times \braket{j_1 (K - \ell) q (-\Delta \ell) | j_1' (K - \ell -\Delta \ell)}^2,
\label{eq:angular_integral_jK}
\end{align}
where $D$ is the Wigner rotation matrix. Importantly, the integral no longer
depends on the values of $\ell$ and $K$ but only on their difference
$K-\ell$, \textit{i.e.}, on the branch under consideration. They also depend
on $\Delta\ell$ which, for a given branch, can be taken as the largest value
allowed by angular-momentum coupling because it corresponds to the largest
radial integral.

Equation~\eqref{eq:angular_integral_jK} describes the change of angular
momentum of the core electron ($j_1 \to j_1'$) through its $q$-pole coupling
with the Rydberg electron. It is particularly instructive because, when multiplied by the radial integrals (see Eq.~\eqref{eq:lopt_contribution_jK_asymp}), it describes an electric dipole ($q=1$) or electric quadrupole ($q=2$) optical transition of the core electron, 
\begin{align}
	&\Gamma^{(q)} \sim 2\pi \left[R_{n \ell}^{\varepsilon \ell', q}D_{0\Delta \ell}^q(0, \frac{\pi}{2}, 0)\right]^2 \nonumber\\
    &\times\left|\braket{N \ell_1 j_1 (K - \ell)| r_1^q  C_{q, -\Delta \ell} | N' \ell_1' j_1' (K - \ell -\Delta \ell)}\right|^2
    \label{eq:asymp_rate_optical_trans}
\end{align}
with an effective ``light'' intensity given by the two terms between the square brackets on the right
hand side. $C_{q, -\Delta \ell}(\theta_1, \phi_1)$ is an unnormalized spherical harmonic.

The transition dipole or quadrupole moment of the core electron in equation~\eqref{eq:asymp_rate_optical_trans}
involves the projection of the core-electron angular momentum $\bm{j}_1$ onto
an axis that is no longer the quantization axis but, rather, another axis
defined by the electron repulsion. The same coefficient also shows that the projection of $\bm{j}_1$ onto the new axis is $K - \ell$.
We have shown earlier that the different branches correspond to different $K-\ell$ values.
We can therefore relate the branches to the orientation of the
core-electron angular momentum relative to the axis defined by its coupling to
the Rydberg electron. Like the branch, this orientation has a crucial influence
on the autoionization of high-$\ell$ core-excited Rydberg states.

For lower values of $\ell$, the angular integrals rapidly change with
$\ell$ (see Fig. \ref{fig:wavefunctions}), and their magnitudes differ
significantly. Transitions that change $K$ become nonnegligible (gray lines in
Fig.~\ref{fig:wavefunctions}). The relative magnitude of the autoionization
rates in different branches can no longer be simply estimated, and we have
observed in all our calculations that the rates become in fact similar
regardless of the branch for values below $\ell \lesssim 10$.

\subsection{General behavior for alkaline-earth species}

\begin{figure}
\centering
\includegraphics[width=\columnwidth]{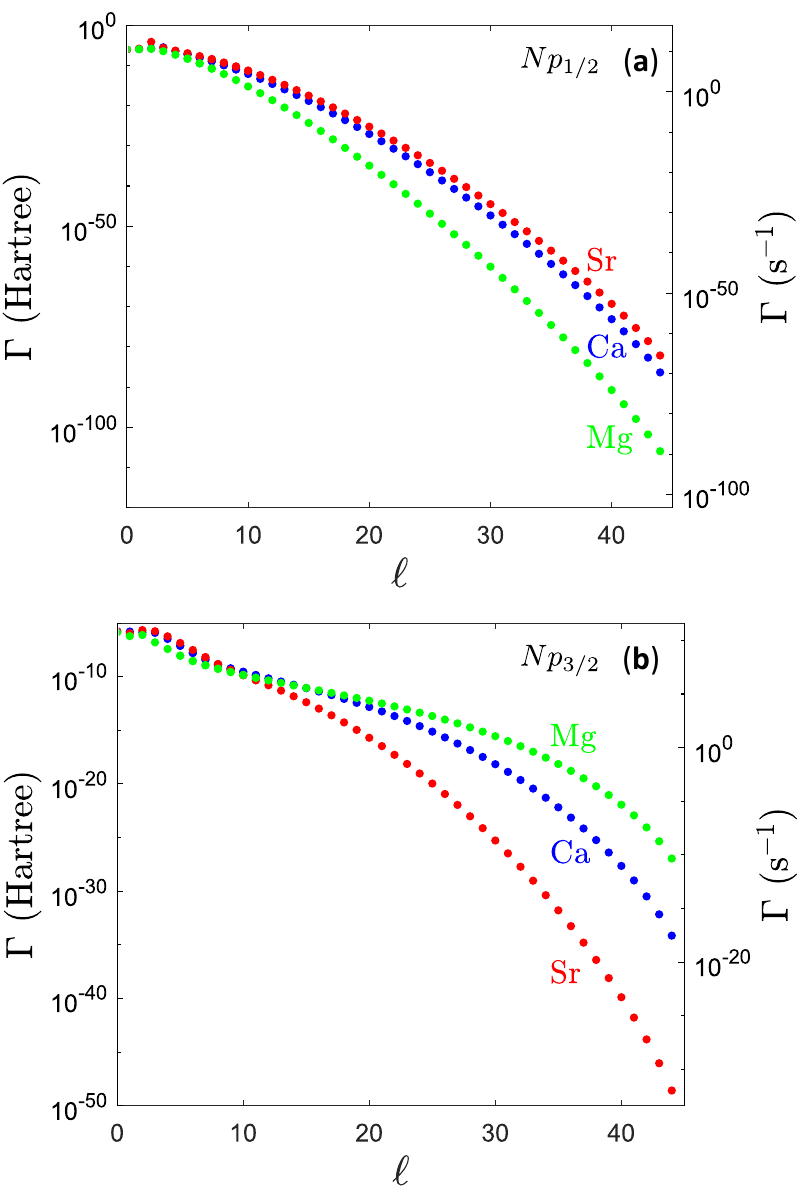}
\caption{\textbf{(a)} Autoionization rates of the $(Np_{1/2}45{\ell}_{j})_J$ states of Mg (solid green circles), Ca (solid blue circles) and Sr (solid red circles) as a function of $\ell$ for the $K - \ell = 1/2$ branch ($j=\ell + 1/2$ and $J=j + 1/2$). \textbf{(b)} Same as panel \textbf{(a)} for the $Np_{3/2}$ thresholds.}
\label{Gamma_l2}
\end{figure}

In addition to Sr, we also calculated the autoionization rates of core-excited
Rydberg states of Mg and Ca. The ionization thresholds $Np_{1/2}, Np_{3/2}$
and, when applicable, $(N-1)d_{3/2}$ and $(N-1)d_{5/2}$, were considered
($N=3-5$, see Fig.~\ref{fig:Energy_schemes}). The rates of the three species
evolve in a similar way with $\ell$, as illustrated by the examples shown in
Fig. \ref{Gamma_l2}. The similarity is not surprising as all
alkaline-earth-metal ions possess the same electronic structure, with the
exception of the $(N-1)d_{3/2,5/2}$ states only present for Ca and the heavier
species. For high $\ell$ values the Rydberg electron is essentially hydrogenic
regardless of the atomic species. Differences in the values of the rates are
thus due to the different properties of the ion cores, in particular the
energies of the states, which affect the photoelectron kinetic energies, and
the transition dipole moment $\left\langle N'\ell_1'j_1'|r_1|N\ell_1j_1
\right\rangle$ and transition quadrupole moment $\left\langle N'\ell_1'j_1'|r_1^2|N\ell_1j_1
\right\rangle$, which directly enter the calculation of the autoionization rates.

For the series converging to the $Np_{1/2}$ thresholds [Fig. \ref{Gamma_l2}(a)],
the continua above the $Ns_{1/2}$ threshold and, except for Mg, the
$(N-1)d_{3/2}$ threshold are accessible. The latter dominate at high $\ell$
values, a fact we verified by calculating the partial rates. The influence of
the photoelectron kinetic energy on the speed at which the rates decrease is
conspicuous. Indeed, the energy difference between the $3s_{1/2}$ and
$3p_{1/2}$ thresholds in Mg (4.422~eV) is much larger than the one between the
$3d_{3/2}$ and $4p_{1/2}$ thresholds in Ca (1.431~eV) or the $4d_{3/2}$ and
$5p_{1/2}$ thresholds in Sr (1.136~eV), and the rates for Mg decay much faster
than for the other two species. We observe that, as before, the faster the
photoelectron the faster the rates decrease with $\ell$.

The rates of series converging to the $Np_{3/2}$ thresholds of Mg, Ca, Sr, and
belonging to the branch with $K - \ell = +\frac{1}{2}$, are shown in Fig.
\ref{Gamma_l2}(b). Autoionization proceeds in the continua above the $Ns_{1/2}$,
$Np_{1/2}$ and, for Ca and Sr, the $(N-1)d_{3/2,5/2}$ ionization thresholds.
At low $\ell$ values, the dipole-type coupling to the $Ns_{1/2}$ and
$(N-1)d_{5/2}$ continua and the quadrupole-type coupling to the $(N-1)d_{3/2}$
continua are all important. We observe, as in Fig.~\ref{fig:Partial_rates_Sr}
and \ref{fig:Branches_Sr(5p32)}, a change in the decay trend around $\ell
\sim 10$ after which autoionization to continua above the $Np_{1/2}$ threshold
is the major decay channel. The same observations as for Fig.
\ref{Gamma_l2}(a)  can be made regarding the relationship between the
photoelectron kinetic energy and the speed at which the rates decrease with
$\ell$. The spin-orbit splitting of the Mg$^+(3p_{1/2, 3/2})$ levels
(11.4~meV) is the smallest of the three species and the rates are the largest
for large $\ell$. The rates for Mg decrease by only 5 orders of magnitude in
the range from $\ell = 10$ to $\ell = 30$, whereas those of Ca and Sr
decrease by 9 and 16 orders of magnitude, respectively. The reasoning holds
for all other branches and all other thresholds of the three species we
investigated.

\subsection{Comparison against available high-$\ell$ experimental data}\label{sec:scaling}

\begin{figure}[h]
\centering
\includegraphics[width=\columnwidth]{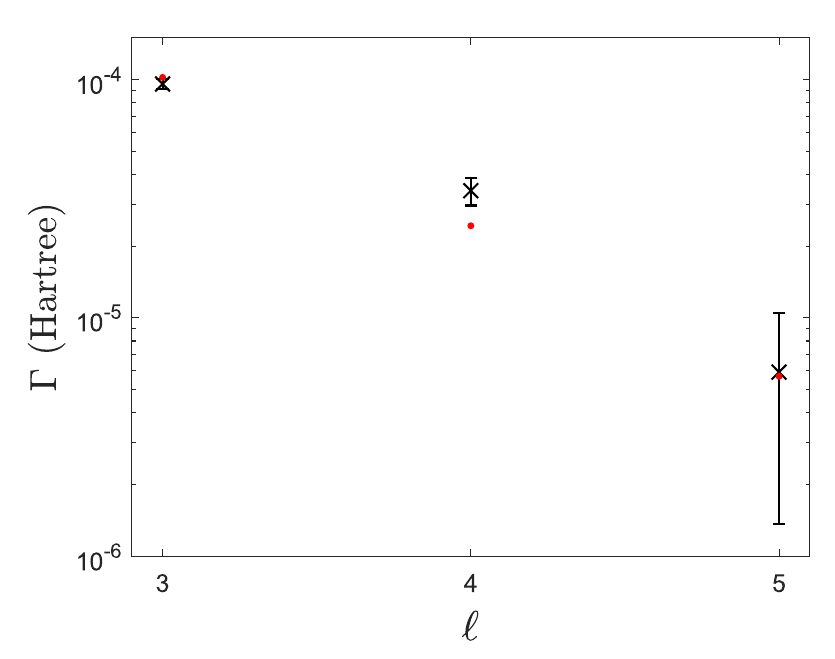}
\caption{Autoionization rates of the $(5p_{1/2}16\ell)$ states of Sr obtained
experimentally (black crosses,~\cite{cooke78a}) and theoretically (red solid
circles, present results). The values shown here are the combined rates for
all the sublevels with same $\ell$ but different $j$ and $J$ values (see
text).}
\label{Gamma_l2_exp}
\end{figure}

Quantitative experimental data on the autoionization rates of high-$\ell$
core-excited Rydberg states are scarce even for the alkaline-earth-metal atoms
which, in comparison, have been extensively studied for low $\ell$ values
(see Ref.~\citenum{aymar96} for a review). Cooke \textit{et
al.}~\cite{cooke78a} measured the autoionization rates of $(5p_{j_1} n
{\ell}_{j})_J$ states of Sr for $n = 16$ and $\ell = 3 \,-\,5$. Their
data, shown in Fig. \ref{Gamma_l2_exp}, falls in good agreement with our
CI-ECS results (black crosses and red solid circles, respectively). In the
experiment, the autoionization rates were determined from the overall
linewidths of $5p_{j_1}n\ell$ states, their $j$ and $J$ substructure
being unresolved. For a given $\ell$ value, the experimental linewidth is
therefore the result of both the combined autoionization linewidths of all
$j$ and $J$ sublevels, and the small energy differences between these
sublevels. We modeled this with our CI-ECS data by generating Lorentzian line
profiles for each sublevel, with center frequencies and linewidths given by
the results of the calculations. These profiles were then summed and the overall
linewidths determined in a least-square fit to a Lorentzian function. It is
these values that are shown in Fig.~\ref{Gamma_l2_exp} (red solid circles).

The $(4d_{3/2, 5/2}51{c}_{j})_J$ circular core-excited Rydberg states of Sr, i.e, those of maximal orbital and magnetic quantum numbers of the Rydberg electron $\left|m\right| = \ell = n-1$,
have been shown to be stable against autoionization by Teixeira \textit{et
al.}~\cite{teixeira20}. They could experimentally determine lower bounds for
the autoionization lifetimes of Sr($4d_{3/2}51c$) and Sr($4d_{5/2}51c$)
circular states of $5$ ms and $2$ ms, respectively. Our calculations agree
with these lower bounds and in fact predict lifetimes that are longer by many
orders of magnitude (77 and 21, respectively). We can thus confirm that
circular core-excited Rydberg states are completely immune to autoionizaton.
Our results reveal that, in fact, most states with $\ell \geq 22$ are also
immune to autoionization, in the sense that autoionization lifetimes are
longer than even the fluorescence lifetime of the Rydberg electron (millisecond range).

In a recent work, Yoshida \textit{et al.}~\cite{yoshida23} thoroughly
investigated the autoionization of the $5p_{1/2}n\ell$ states of Sr with $\ell =
0-5$. To compare our CI-ECS results with the data presented
in~\cite{yoshida23}, we determined the scaled autoionization rates $\Gamma_0$
of the $(5p_{1/2}n\ell_j)_J$ series ($\ell=3-5$) by fitting the calculated
rates to the usual formula
\begin{equation}
	\Gamma(n) = \frac{\Gamma_0}{n^3}
	\label{eq:rates_scaling_law}
\end{equation}
in the range $n=51-75$. The scaled rates are compared in
Table~\ref{tab:comparison_yoshida} and show good agreement with the results
of~\cite{yoshida23}. For lower $\ell$ values, the rates do not closely follow
the scaling law~\eqref{eq:rates_scaling_law} because of series perturbations.

\begin{table}
	\centering
	\begin{ruledtabular}
	\caption{Comparison between the experimental and theoretical scaled autoionization rates $\Gamma_0$ from Ref.~\cite{yoshida23} and the present theoretical values for the ($5p_{1/2}n\ell_j)_J$ Rydberg series of Sr ($\ell=3-5$). In the experiment~\cite{yoshida23}, the rates of Rydberg series with the same $\ell$ values but different $j$ and $J$ values could not be individually determined.}
	\label{tab:comparison_yoshida}
	\begin{tabular}{clll}
	Series & $\Gamma_0^\text{exp}$~\cite{yoshida23} & $\Gamma_0^\text{th}$~\cite{yoshida23} & $\Gamma_0^\text{th}$~(present) \\\hline
	$(5p_{1/2}nf_{5/2})_{J=2}$  &  \multirow{4}{*}{0.181} & 0.199 & 0.234 \\
	$(5p_{1/2}nf_{5/2})_{J=3}$  &   & 0.164 & 0.165 \\
	$(5p_{1/2}nf_{7/2})_{J=3}$  &   & 0.201 & 0.207 \\
	$(5p_{1/2}nf_{7/2})_{J=4}$  &   & 0.230 & 0.194 \\[0.2cm]
	$(5p_{1/2}ng_{7/2})_{J=3}$  &  \multirow{4}{*}{0.056} & 0.081 & 0.069 \\
	$(5p_{1/2}ng_{7/2})_{J=4}$  &   & 0.046 & 0.042 \\
	$(5p_{1/2}ng_{9/2})_{J=4}$  &   & 0.081 & 0.069 \\
	$(5p_{1/2}ng_{9/2})_{J=5}$  &   & 0.046 & 0.042 \\[0.2cm]
	$(5p_{1/2}nh_{9/2})_{J=4}$  &  \multirow{4}{*}{0.027} & 0.015 & 0.018 \\
	$(5p_{1/2}nh_{9/2})_{J=5}$  &   & 0.006 & 0.009\\
	$(5p_{1/2}nh_{11/2})_{J=5}$ &   & 0.015 & 0.018 \\
	$(5p_{1/2}nh_{11/2})_{J=6}$ &   & 0.006 & 0.010
	\end{tabular}
	\end{ruledtabular}
\end{table}

\subsection{$\ell$ and $n$ scaling}

\begin{figure}[ht]
    \centering
	\includegraphics[width=\columnwidth]{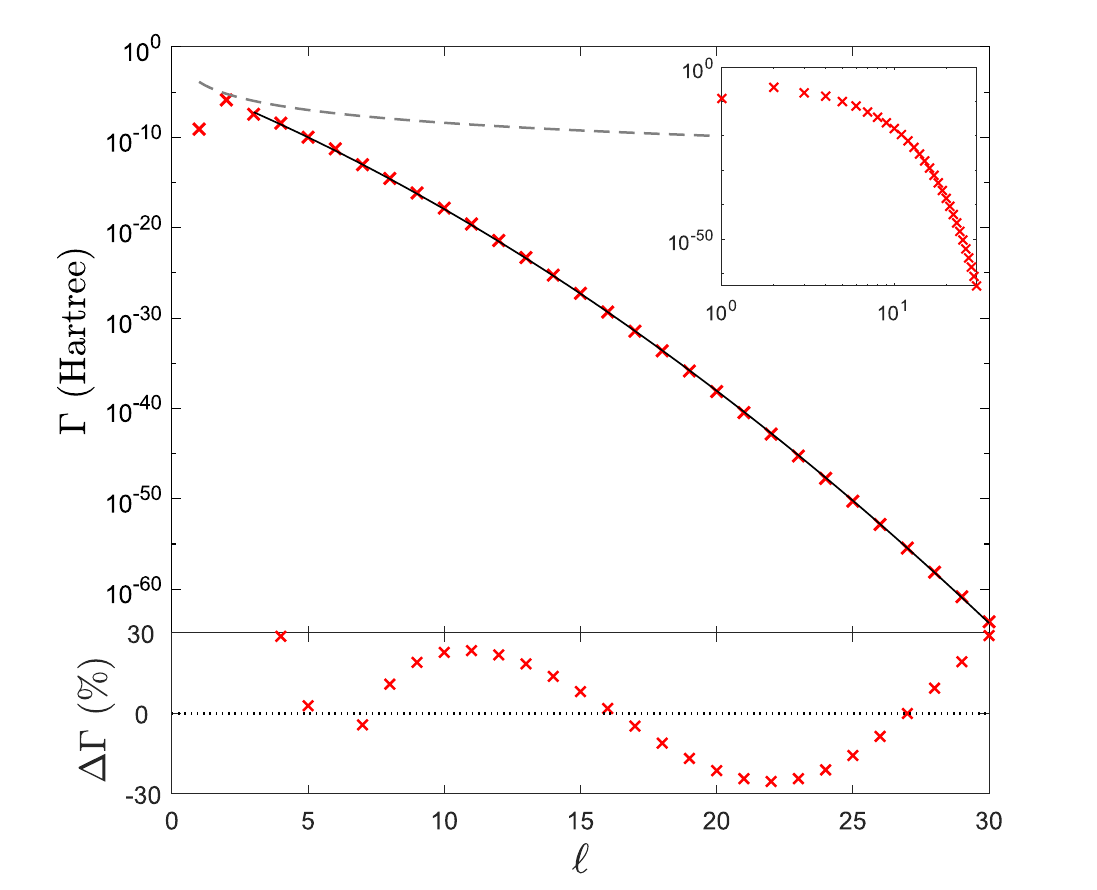}
    \caption{Calculated autoionization rates (red crosses) of the $(3p_{1/2}45{\ell}_{j})_J$ states of Mg for the $K-\ell = -\frac{1}{2}$ branch ($j = \ell - \frac{1}{2}$ and $J = j - \frac{1}{2}$). Formula~\eqref{eq:AI_rates_1comp} is shown by the black solid line, with parameters obtained in a least-squares fit of the calculated data in the range $\ell = 5 - 30$. The gray dashed line corresponds to the polynomial scaling discussed in text. The inset shows the same values on a log-log scale, highlighting the absence of polynomial scaling of the rates with $\ell$. The bottom part of the graph shows the fit residuals on a relative scale.}
    \label{fig:fit_mg}
\end{figure}

Beyond $\ell \sim 4$ Rydberg-series perturbations are rare and the behavior of the autoionization rates with $\ell$ is smooth. We found that the decrease of the rates with $\ell$ is well described by the empirical exponential law
\begin{equation}
\Gamma(\ell) = \Gamma_0 e^{a\ell^2 + b\ell} ,
\label{eq:AI_rates_1comp}
\end{equation}
where $\Gamma_0$, $a$ and $b$ are parameters that depend on $n$ and on the
branch under consideration. They can be determined in a least-squares fit of
the calculated rates ($\ell = 5 - 30$) yielding, for example, $\Gamma_0 =
3.5(15) \cdot 10^{-4}$ Hartree, $a = -0.064(2)$ and $b = -2.69(6)$ for the
$(3p_{1/2}45{\ell}_{\ell - 1/2})_{j - 1/2}$ series of Mg. The law is
shown by the solid black line in Fig.~\ref{fig:fit_mg} and compared against the
calculated data (red crosses). It reproduces the rates to within $30\%$ or
better over the 60 orders of magnitude that their values span. Increasing the
degree of the polynomial in the exponential leads to a more accurate fit, at
the expense of an increased number of parameters. We find that a second order
polynomial represents a good compromise between accuracy and simplicity.

\begin{figure}
    \centering
	\includegraphics[width=\columnwidth]{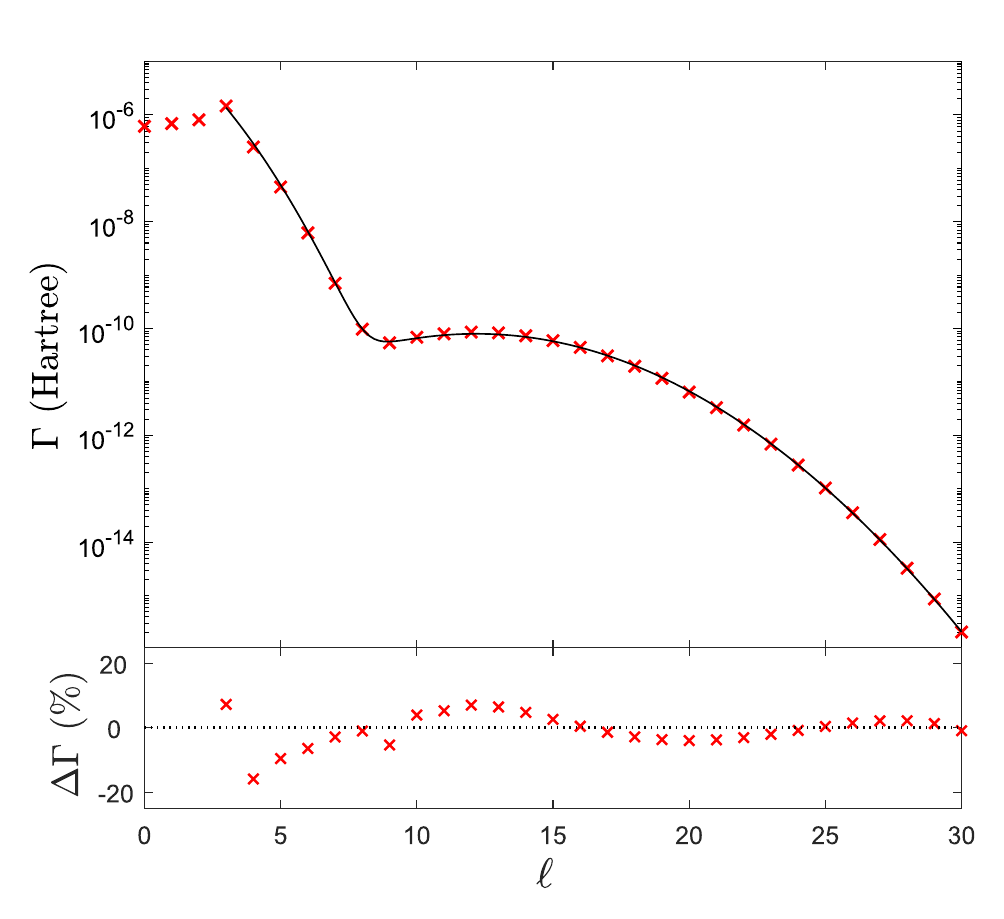}
    \caption{Calculated autoionization rates of the Ca$(4p_{3/2}45{\ell}_{j})_J$ series with $j = \ell + 0.5$ and $J = j + 1.5$ (red crosses). The solid line shows formula~\eqref{AI_rates_2comp} with $k=2$ and parameters obtained in a least-squares fit of the calculated data in the range $\ell = 5 - 30$. The bottom part of the graph shows the fit residuals on a relative scale.}
    \label{Fit_Ca}
\end{figure}

Equation~\eqref{eq:AI_rates_1comp} describes the rates well when a single
decay trend is observable. When the rates show several decay trends, typically
associated with autoionization into the continua of different ion-core states,
the behavior is well described by the sum of exponential laws
\begin{equation}
\Gamma(\ell) = \sum_{i=1}^k \Gamma_0^{(i)} e^{a^{(i)}\ell^2 + b^{(i)}\ell} \;\;\; .
\label{AI_rates_2comp}
\end{equation}
$k$ is typically the number of ion-core thresholds with significantly
different energies. The rate values obtained by fitting the above equation,
choosing $k=2$, to the theoretical results for the Ca$(4p_{3/2}45\ell_{\ell+1/2})_{\ell+2}$ branch ($\ell = 5 - 30$)
are shown in Figure~\ref{Fit_Ca}. As for the single-trend case, the agreement
between the fit results and the calculated values is excellent and better than
$20\%$ over the entire range ($\ell = 5-30$).

\begin{figure}
\centering
\includegraphics[width=\columnwidth]{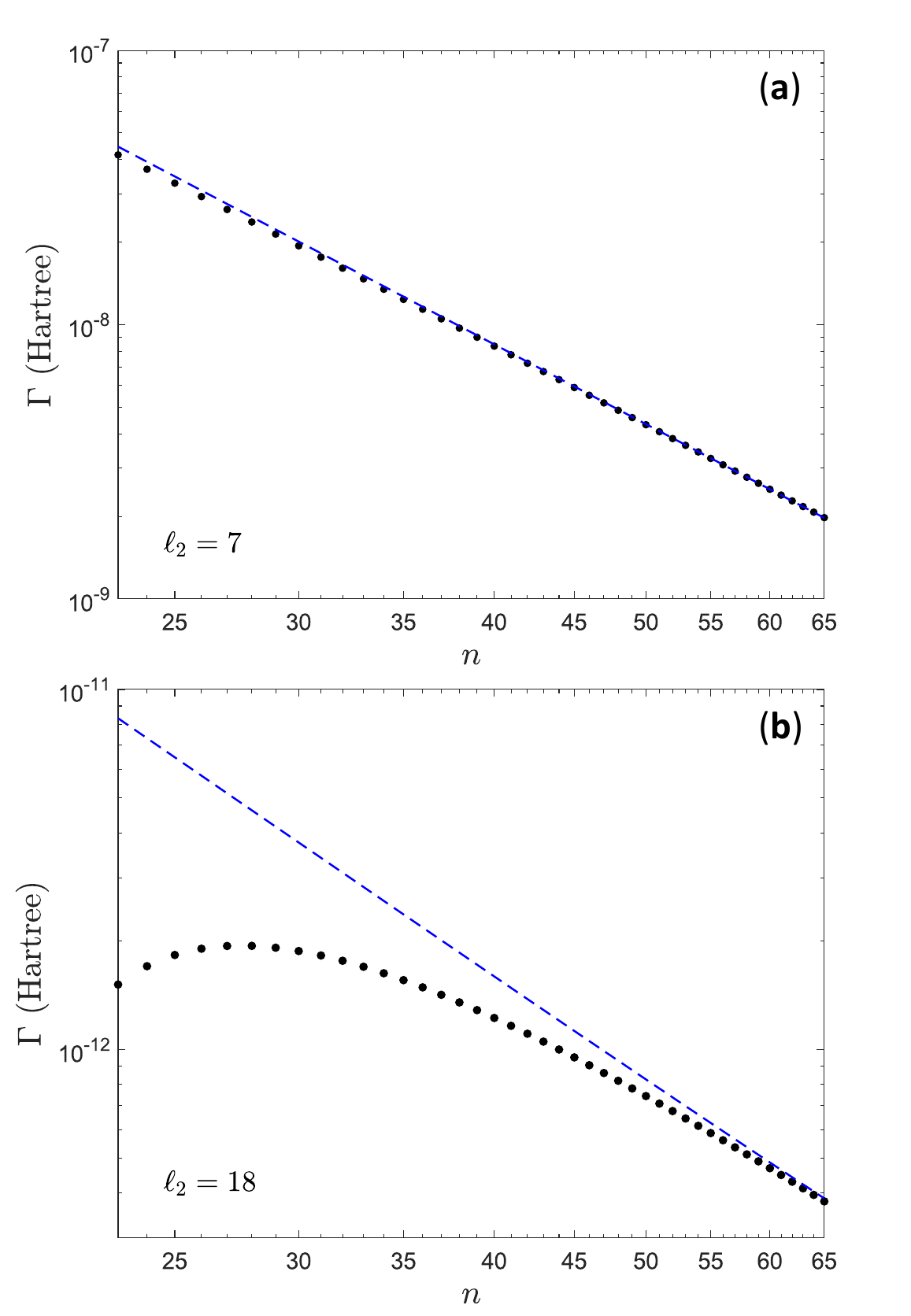}
\caption{Dependence of the calculated autoionization rates on $n$ for the $(4p_{3/2}n\ell_{j})_J$ Rydberg series of Ca with \textbf{(a)} $\ell = 7$, and \textbf{(b)} $\ell = 18$. In both cases we chose $j = \ell + 1/2$ and $J = j - 1/2$. The blue dashed line represents the $n^{-3}$ scaling law reached when $n \gg \ell$. For $n < 23$, autoionization into the continua above the Ca$^+(4p_{1/2})$ threshold is no longer energetically allowed and the rates are much smaller.}
\label{Gamma_n2}
\end{figure}

An approximate formula for the $\ell$ dependence of autoionization rates,
derived in Ref.~\cite{cooke79}, predicts a polynomial dependence $\Gamma
\propto 1/n^3\ell^{4q-3}$ where $q$ is the order of the
multipole expansion giving the dominant contribution to autoionization. The
prediction has been verified for the low $\ell$ states of the
Sr($5p_{1/2}n\ell$) series~\cite{yoshida23}  ($\ell \lesssim 5$) but, as $\ell$
increases, it rapidly deviates from our results. The same is true for other
series and species and, in the case of Mg shown in Fig.~\ref{fig:fit_mg}
(dashed gray line), the deviation occurs at even smaller $\ell$ values. The
deviation can be attributed to the assumptions made in Ref.~\cite{cooke79} to analytically estimate
the radial integral~\eqref{eq:radial_rydberg} and the
rate~\eqref{eq:golden_rule}, leading to the polynomial scaling, whereas the
integral is calculated exactly in the present work.

The $n^{-3}$ scaling of autoionization rates is well established for low $\ell$
values but deserves a closer inspection as $\ell$ increases. When $n \gg
\ell$, the cubic scaling is verified~\cite{gallagher94} as illustrated in
Fig. \ref{Gamma_n2}(a) for the $(4p_{3/2}n(\ell=7)_{15/2})_{7}$ series of
Ca. Clear deviations appear when $n$ becomes comparable to $\ell$, as
shown in Fig. \ref{Gamma_n2}(b) for the same branch but a higher $\ell=18$
value. In this situation, the rates initially increase with $n$ before
passing through a maximum and eventually
following the expected $n^{-3}$ asymptotic form (blue dashed line) as $n \gg
\ell$. The same observation holds for all other branches, ion-core states,
and atomic species that we studied.

A departure from the $n^{-3}$ scaling law is well known for the fluorescence
lifetimes of high-$\ell$ Rydberg states. In this case, because the
fluorescence only occurs to nearby states of similar $n$ values, a scaling of
$n^{-5}$ can be derived~\cite{gallagher94}. Autoionization proceeds, instead,
to continua with the same energy regardless of the value of $\ell$ and, therefore, an
argument similar to the one for fluorescence cannot be made. The $n$
dependence of the autoionization rates, encoded in the complicated functions shown in Eq.~\eqref{eq:radial_rydberg}, depends on the
kinetic energy of the ionized electron and does not follow a simple polynomial
scaling law.

\section{Conclusion}

The autoionization rates of core-excited states of Mg, Ca and Sr were
calculated for $n = 10 - 65$ and $\ell = 0 - 45$. For low $\ell$ values,
we obtained the rates by treating the full extent of correlations between the
two valence electrons with CI-ECS. Both the values of the rates and the
perturbations caused by states belonging to other, adjacent Rydberg series
fall in excellent agreement with the available experimental data. Beyond
$\ell \sim 5$, the rates drop rapidly and perturbations become much scarcer,
two facts indicating the rapid decrease of the electron-electron repulsion. We
show that a perturbative treatment of dipole- and quadrupole-type electron correlations, which compares
well to the results of the full nonperturbative CI-ECS calculations, is
sufficient at this point. The hydrogenic integrals involved in the
perturbative calculations are computed without approximation and with a
numerical precision of $10^{-60}$, a fact imposed by the rapid decrease of the
rates with $\ell$.

The complete picture provided by the results has allowed us to analyze the
autoionization of high-$\ell$ core-excited Rydberg states in detail and
derive both the quantum number dependencies of their autoionization dynamics and the physical mechanisms responsible for these dependencies. Five general laws have been identified:

First, the decay of the autoionization rates with $\ell$ is very rapid. Above
$\ell \sim 25$, they become negligible compared to all other decay processes,
even those as slow as the fluorescence of the Rydberg electron taking place on
the millisecond timescale.

Second, the values of the rates separate into branches belonging to different
$K-\ell$ values, \textit{i.e.}, different couplings between the total
angular momentum of the ion core and the orbital angular momentum of the
Rydberg electron. Each branch can be associated to predominantly one change of
$\ell$ upon autoionization ($\Delta \ell = \pm 1$ for $q=1$ and $\Delta
\ell = 0, \pm 2$ for $q=2$), a property resembling selection rules in
radiative transitions. Depending on the predominant $\Delta \ell$, the
values of the rates differ by up to several orders of magnitude which gives
rise to well separated branches.

Third, for each branch the decrease of the rates with $\ell$ presents a single
decay trend if autoionization proceeds predominantly into the continua above a
single ion-core state. Otherwise, several decay trends can be observed. The
speed at which the rates decrease is determined by the energy of the
autoionized electron, therefore the different trends are particularly
pronounced when autoionization occurs into continua above ion-core states
that have very different energies. In that case, a shoulder is observed around $\ell
\sim 8$ where the rapid decrease of the rates suddenly turns into a much
slower one. A similar trend has been observed in other species such as the Yb
atom~\cite{lehec21}.

Fourth, autoionization rates are typically larger when the kinetic energy of
the electron is small, and smaller when this energy is large. This means, for
example, that the rates for the Mg($3p_{1/2}n\ell$) series are smaller
than those of the Ca($4p_{1/2}n\ell$) and Sr($5p_{1/2}n\ell$) series
because the Mg$^+(3p_{1/2})$ ion-core state lies the highest in energy.

Fifth, the autoionization rates decrease with $\ell$ following, to a good
approximation, an exponential law in which the argument is a second-order
polynomial in $\ell$. Using this law the rates can be described within a
good relative accuracy over the many orders of magnitude that they span. This
scaling law can be used, in the future, to extrapolate the rates from a small
set of measured high-$\ell$ autoionization rates to other $\ell$ values.
The dependence of the rates on $n$ follows the usual $n^{-3}$ scaling
law when $n \gg \ell$. This is no longer the case when $n$ and $\ell$
are similar, in which case no simple scaling law has been found.

The general laws presented above were derived from the extensive data
calculated for the alkaline-earth-metal atoms. For high-$\ell$ states, the
exact shape of the ion core has only little influence on the Rydberg electron
as, because of the centrifugal barrier, it does not penetrate in the ion core
region. The conclusions drawn above are therefore not limited to
alkaline-earth-metal species and are expected to apply to high-$\ell$
Rydberg states of other atoms, molecules and ions. Generalization to the case
of molecules requires the vibrational and rotational structure of the ion core
to be taken into account, leading to several differences compared to the
atomic case. More branches are expected to form because the electronic angular
momenta also couple to the rotational angular momentum of the ion core
following, typically, Hund's angular-momentum-coupling cases (d) or
(e)~\cite{lefebvre-brion04b}. Rotational and vibrational autoionization can
occur and typically involve small energies for the ionized electron. If these
decay channels dominate, we expect that the decay of the autoionization rates
with $\ell$ be significantly slower. A comprehensive study of the
autoionization of the high-$\ell$ Rydberg states of molecules is
an interesting perspective for future work.

\begin{acknowledgements}
	We would like to thank F.\,Merkt for bringing the subject of the present paper to our attention, and X.\,Urbain for helpful comments. This work was supported by the Fonds de la Recherche Scientifique (FNRS) under IISN Grant No.~4.4504.10. E.M.B. and M.G. acknowledge support from the Fonds Spéciaux de Recherche (FSR) of UCLouvain. Computational resources have been provided by the supercomputing facilities of the Université catholique de Louvain (CISM/UCL) and the Consortium des Équipements de Calcul Intensif en Fédération Wallonie Bruxelles (CÉCI) funded by the Fond de la Recherche Scientifique de Belgique (F.R.S.-FNRS) under convention 2.5020.11 and by the Walloon Region. 
\end{acknowledgements}

\appendix
\section{Asymptotic formula for the angular integrals in $jK$ coupling}\label{sec:asymp_jk_coupling}

In the perturbative limit and neglecting exchange, each multipole contribution $\Gamma^{(q)}$ to the total autoionization rate can be written in the $jK$ coupling scheme as (see, \textit{e.g.}, Refs.~\cite{poirier88,poirier94a} for details)
\begin{widetext}
\begin{equation}
	\Gamma^{(q)} = 2\pi \left[R_{N \ell_1 j_1}^{N' \ell'_1 j'_1, q}R_{n \ell}^{\varepsilon \ell', q}\right]^2
    [\ell_1, \ell_1', \ell, \ell', j_1, j_1']
    \begin{pmatrix}
		\ell_1' & q & \ell_1 \\
		0    & 0 & 0
	\end{pmatrix}^2
	\begin{pmatrix}
		\ell' & q & \ell \\
		0    & 0 & 0
	\end{pmatrix}^2
	\begin{Bmatrix}
		j_1  & q & j'_1 \\
		\ell_1' & 1/2 & \ell_1
	\end{Bmatrix}^2
	\begin{Bmatrix}
		\ell'  & j_1' & K \\
		j_1 & \ell & q
	\end{Bmatrix}^2 .
	\label{eq:lopt_contribution_jK}
\end{equation}
\end{widetext}
The symbols are defined as in Eq.~\eqref{eq:lopt_rate}. The limit of large $\ell$ values implies that $K$ is large because $j_1$ is small. Because $q$ is small, $\ell'$ is large and $j_1'$ is small. Using the symmetry of the $6j$ symbols and the asymptotic formula given in Ref.~\cite{varshalovich88b}, the last squared Wigner $6j$ symbol in the above equation simplifies to
\begin{equation}
	\begin{Bmatrix}
		\ell'  & j_1' & K \\
		j_1 & \ell & q
	\end{Bmatrix}^2
	\simeq \frac{\braket{j_1 b j_1' (\Delta \ell - b) | q \Delta \ell}^2}{2\ell(2q+1)} .
\end{equation}
We defined $b = K-\ell$, which labels in fact the autoionization branch (see Sec.~\ref{sec:results_allbranches}). The quantity $\Delta \ell = \ell' - \ell$ represents the change in orbital angular momentum of the Rydberg electron upon autoionization. Using the symmetry properties of Clebsch-Gordan coefficients one can rewrite the above equation as
\begin{equation}
	\begin{Bmatrix}
		\ell'  & j_1' & K \\
		j_1 & \ell & q
	\end{Bmatrix}^2
	\simeq \frac{\braket{j_1 b q (-\Delta \ell) | j_1' (b-\Delta \ell)}^2}{2\ell(2j_1'+1)} .
	\label{eq:asymptotic_expression_6j}
\end{equation}
The square of the $3j$ symbol involving $\ell$ and $\ell'$ in Eq.~\eqref{eq:lopt_contribution_jK} can also be simplified in the limit $\ell, \ell' \gg q$. Using the asymptotic expression for Clebsch-Gordan coefficients of Ref.~\cite{varshalovich88c}, we have
\begin{equation}
	\begin{pmatrix}
		\ell' & q & \ell \\
		0    & 0 & 0
	\end{pmatrix}^2
	\simeq
	\frac{\left[D_{0\Delta \ell}^q(0, \frac{\pi}{2}, 0)\right]^2}{2\ell'+1} ,
	\label{eq:asymptotic_expression_3j}
\end{equation}
where $D_{0\Delta \ell}^q(0, \frac{\pi}{2}, 0)$ is the Wigner $D$ matrix. Replacing Eqs.~\eqref{eq:asymptotic_expression_6j} and~\eqref{eq:asymptotic_expression_3j} in Eq.~\eqref{eq:lopt_contribution_jK}, one obtains
\begin{widetext}
\begin{equation}
	\Gamma^{(q)} = 2\pi \left[R_{N \ell_1 j_1}^{N' \ell'_1 j'_1, q}R_{n \ell}^{\varepsilon \ell', q}\right]^2
    [\ell_1, \ell_1', j_1]
    \begin{pmatrix}
		\ell_1' & q & \ell_1 \\
		0    & 0 & 0
	\end{pmatrix}^2
	\begin{Bmatrix}
		j_1  & q & j'_1 \\
		\ell_1' & 1/2 & \ell_1
	\end{Bmatrix}^2
	\braket{j_1 b q (-\Delta \ell) | j_1' (b-\Delta \ell)}^2
	\left[D_{0\Delta \ell}^q(0, \frac{\pi}{2}, 0)\right]^2
	,
	\label{eq:lopt_contribution_jK_asymp}
\end{equation}
\end{widetext}
where we used $2\ell + 1 \simeq 2\ell$. The angular part of this equation,
\textit{i.e.}, the terms on the right hand side that depend on the
angular-momentum quantum numbers, is the one given in
Eq.~\eqref{eq:angular_integral_jK}.

\bibliographystyle{apsrev4-2}

\end{document}